\documentclass[lettersize,journal]{IEEEtran}
\usepackage{amsmath,amsfonts}
\usepackage[T1]{fontenc}
\usepackage{algorithm}
\usepackage{cite}
\usepackage{algorithmic}
\usepackage{array}
\usepackage{threeparttable}
\usepackage{multirow}
\usepackage{bm}
\usepackage[justification=centering]{caption}
\usepackage[caption=false,font=normalsize,labelfont=sf,textfont=sf]{subfig}
\usepackage{textcomp}
\usepackage{stfloats}
\usepackage{tabularx}
\usepackage{url}
\usepackage{verbatim}
\usepackage{graphicx}
\usepackage{makecell}
\usepackage{float}
\usepackage[justification=raggedright,singlelinecheck=false]{caption}
\def\BibTeX{{\rm B\kern-.05em{\sc i\kern-.025em b}\kern-.08em
    T\kern-.1667em\lower.7ex\hbox{E}\kern-.125emX}}
\usepackage{balance}
\renewcommand{\vec}[1]{\bm{\mathrm{#1}}} 
\begin{document}

\title{Grid-Reg: Detector-Free Gridized Feature Learning and Matching for Large-Scale SAR-Optical Image Registration}

\author{\IEEEauthorblockN{Xiaochen Wei, Weiwei Guo,~\IEEEmembership{Member,~IEEE}, Zenghui Zhang,~\IEEEmembership{Senior Member,~IEEE},~Wenxian Yu,~\IEEEmembership{Senior Member,~IEEE}}

\thanks{This work was supported by the National Natural Science Foundation of China  62071333. (\emph{Corresponding author: Weiwei Guo})
 Xiaochen Wei, Wenxian Yu are with Shanghai Key Laboratory of Intelligent Sensing and Recognition, Shanghai Jiao Tong University, Shanghai 200240, China.
 Weiwei Guo is with Center of Digital Innovation, Tongji University, Shanghai 200092, China.(email: weiweiguo@tongji.edu.cn)  
 }}


\maketitle
\begin{abstract}

It is highly challenging to register large-scale, heterogeneous SAR and optical images, particularly across platforms,  due to significant geometric, radiometric, and temporal differences, which most existing methods struggle to address. To overcome these challenges, we propose Grid-Reg, a grid-based multimodal registration framework comprising a domain-robust descriptor extraction network, Hybrid Siamese Correlation Metric Learning Network (HSCMLNet), and a grid-based solver (Grid-Solver) for transformation parameter estimation. In heterogeneous imagery with large modality gaps and geometric differences, obtaining accurate correspondences is inherently difficult. To robustly measure similarity between gridded patches, HSCMLNet integrates a hybrid Siamese module with a correlation metric learning module (CMLModule) based on equiangular unit basis vectors (EUBVs), together with a manifold consistency loss to promote modality-invariant, discriminative feature learning. The Grid-Solver estimates transformation parameters by minimizing a global grid matching loss through a progressive dual-loop search strategy to reliably find patch correspondences across entire images. Furthermore, we curate a challenging benchmark dataset for SAR-to-optical registration using real-world UAV MiniSAR data and Google Earth optical imagery. Extensive experiments demonstrate that our proposed approach achieves superior performance over state-of-the-art methods.

\end{abstract}
\begin{IEEEkeywords}
Multimodal Image Registration, Detector-free, SAR, Feature Learning
\end{IEEEkeywords}
\section{Introduction}
\IEEEPARstart{S}aceborne and ariborne synthetic Aperture Radar (SAR) systems have become indispensable sensors for diverse Earth observation. Compared to optical sensors, SAR images reflect electromagnetic scattering characteristics of the object surfaces, which are quite difficult for human visual perception\cite{Mihai2023}. Usually, corresponding optical images are employed to provide complementary visual cues, facilitating the joint interpretation of SAR and optical images. Consequently, precise SAR-optical image registration is a fundamental prerequisite for the joint interpretation, as well as for a broad range of downstream tasks, including data fusion, geo-localization, change detection, and object recognition\cite{Zhang2025,Zhu2023}, and to name just a few. Although numerous methods have been proposed in recent decades, from traditional feature-based to modern deep learning-based approaches, SAR-Optical image registration remains highly challenging, struggling against significant radiometric, geometric differences and temporal discrepancies. 

\begin{figure}[!htbp]
	\includegraphics[width=0.9\linewidth]{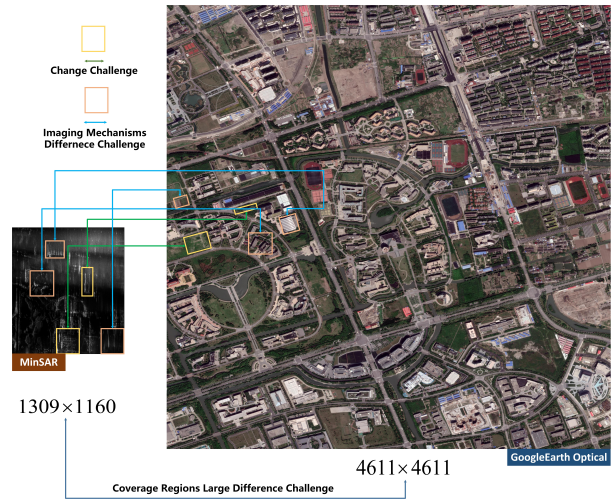}
	\caption{{Significant geometric, radiometric and temporal  differences of UAV SAR and Satellite Optical Images from Google Earth.}}
	\label{factors}
\end{figure}

Most existing methods for multimodal remote sensing image registration \cite{44,70,76} achieve good performance only on small-sized tiled image pairs of size typically less than one thousand pixels in each dimension, with limited geometric and radiometric differences. However, as illustrated in Figure~\ref{factors}, real-world SAR and optical remote sensing images, especially airborne SAR and spaceborne optical images, are often extremely large, exceeding several thousand pixels per side, exhibiting substantial geometric and radiometric differences, and even temporal variations. Current methods fail to handle such challenging scenarios. On one hand, deep learning-based approaches are seldom directly applied to such large-scale images, not only due to GPU memory limitations but also the prohibitive computational cost and lower accuracy. On the other hand, feature-based matching approaches are usually adopted, which rely on accurate keypoint correspondences across images to estimate the registration transformation. However, establishing reliable and robust feature correspondences between large-scale, multimodal SAR and optical images remains extremely challenging, as shown in Figure~\ref{factors}, which lies in:

\begin{enumerate}
\item Firstly, the appearances of SAR and optical images are completely distinct due to their essentially different imaging mechanisms, making it very difficult to identify common features for reliable matching. Moreover, airborne SAR images often suffer from defocusing, speckle noise, and geometric distortions, further hindering the extraction of robust and consistent features. 
\item Secondly, airborne or UAV SAR and satellite optical images acquired across different platforms often exhibit large geometric discrepancies, including translations and rotations. The position and orientation system (POS) data of airborne platforms may be imprecise or even unavailable in GNSS-denied environments, which limits the ability of geographical calibration for airborne or UAV SAR data, compared to satellite systems with more stable and accurate georeferencing. The geometric differences are significantly greater than those observed in same-platform scenarios, where registration is relatively easier.   
\item Thirdly, SAR and optical images are usually captured at different times, resulting in the possible temporal changes in land cover and land use. For example, in Fig. \ref{factors},  new buildings presented in SAR images are absent in the corresponding optical image.
\end{enumerate}

In this paper, we aim to develop a multimodal registration method for more practical, large-scale SAR and optical images acquired across different platforms, which exhibit significant geometric and radiometric differences. This problem remains largely unexplored in the literature. We propose a novel grid-based registration framework, Grid-Reg, for such challenging large-scale, heterogeneous remote sensing image registration across platforms. Our proposed method is detector-free, identifying patch correspondences between airborne SAR and optical images and estimating transformation parameters by optimizing a global grid-matching objective. 

It is then crucial to learn discriminative and modality-invariant patch descriptors to enable robust matching between SAR and optical images in the presence of substantial modality, geometric and temporal discrepancies. Current methods mainly adopt metric learning techniques based on the Siamese network architecture, where fully connected layers are deployed on top of the network to extract high-dimensional feature representations. The contrastive-like metric learning loss is usually minimized to pull samples from the same categories closer while pushing the samples from different categories farther apart. Although originally designed for classification tasks, they are not well-suited for the registration task that aims to learn salient, discriminative and modality-invariant features between cross-modal image pairs. In particular, these methods are vulnerable to noisy or temporally changing input pairs, resulting in inconsistent and unstable matching results. In this paper, we proposed a manifold-awared cross-modal feature learning method based on Equiangular Unit basis Vectors (EUBVs) representation that learns the cross-modal patch embeddings by the deep neural network onto a normalized hypersphere with Equiangular Unit Vectors as basis. The cross-modal consistency loss is designed to reconstruct the shared basis (e.g., EUBvs) using the representation coefficients from one modality together with the feature representations from the other modality, and vice versa, encouraging the extraction of modality-invariant features.

Furthermore, we propose a detection-free, grid-based estimator for registration parameters that adopts a progressive strategy to optimize a novel global grid-level feature matching objective, enabling robust and accurate registration across large-scale heterogeneous SAR and optical images.

Due to the lack of such a challenging benchmark dataset for large-scale SAR and optical image registration, we curate a new benchmark dataset utlizing UAV MiniSAR imagery and the corresponding Google Earth optical images to evaluate the capability of the cross-modal registration methods.
$FN$

Our contributions are summarized as follows:
\begin{itemize}
\item We propose a grid-based registration framework, \textbf{Grid-Reg}, for large-scale SAR and optical image registration across platforms with significant appearance and geometric variations, avoiding fragile keypoint detection and matching. To the best of our knowledge, this is the first work to address the registration of large-scale airborne SAR and spaceborne optical images with extreme modality gap and differences..
\item We design a novel multimodal patch descriptor network (\textbf{HSCMLNet}) for matching, which integrates a hybrid Siamese structure and a modality-invariant metric learning module based on Equiangular unit basis representations. The proposed network is capable of extracting discriminative and modality-invariant features from SAR and optical images, being robust against significant modality gaps, noise, and temporal variations. A modality-consistency loss is introduced to encourage modality-invariant feature learning. 
\item We introduce a grid-based registration estimation method, \textbf{Grid-Solver}, which progressively minimizes a novel grid matching objective in a dual-loop search strategy. This strategy iteratively explores grid point correspondences in an outer loop and refines them within an expanded candidate set in the inner loop, effectively avoiding local optima and achieving the minimal global matching loss. Compared with feature-based methods that are susceptible to a large number of mismatched point pairs, Grid-Solver enables robust parameter estimation by leveraging patch-level grid correspondences and triangle-based geometric constraints. 

\item We develop a new benchmark dataset consisting of our UAV-based MiniSAR  and  Google Earth optical satellite images of large size, significant modality and geometric differences, and temporal changes. We conduct extensive experimental analysis and comparison, demonstrating the effectiveness of our proposed registration approach.
\end{itemize}


\section{Related Work}
Multi-modal Image registration methods can be categorized into feature-based, area-based, and deep learning methods \cite{35}.
\subsection{Feature-based Image Registration}
Feature-based methods usually involve three key steps: key-point detection, descriptor extraction and matching, and robust transformation parameter estimation. It is desired that the salient key points are invariant to modality, geometry, and perspective variations. Traditional methods rely on manually crafted features like SIFT, Harris, and SURF. Recent Deep neural networks \cite{10,11,12,49} dominate learning local descriptors for image registration, surpassing traditional methods.  Nina et.al \cite{75} used modified HardNet features to align SAR and optical images. Cnet \cite{76} introduced a feature learning network with convolutional layers and an attention module to enhance feature representation. To retain discriminative information while eliminating speckle noise in SAR images, Xiang et.al\cite{25} proposed a feature decoupling network (FDNet) that combined a residual denoising network and a pseudo-siamese fully convolutional network. To handle mismatches for parameter estimation, some robust estimation algorithms, for example RANSAC, GMS \cite{30}, VFC \cite{Ma2014RobustPM} were proposed. SuperGlue \cite{31} simultaneously matches features and removes outliers using a graph neural network-based loss function. Zhao et al. \cite{33} adopted a gravity centre coordinate system for geometrically invariant keypoint detection, achieving high-quality matching pairs with their SMFNet feature matching network. To estimate transformation parameters, Brian et al. \cite{55} proposed a method to estimate global transformation parameters by first estimating local geometric transformations to align large-scale images. Cédric et al. \cite{71} adopted entropy and mutual information to estimate parameters for low-overlap image registration.

The success of local feature-based methods largely relies on high-quality modality-invariant descriptor for matching. However, current methods usually adopt metric learning techniques to learn holistic descriptors that aim to maximize the similarity between positive pairs while minimizing it between negative ones. These methods are vulnerable to significant geometric, appearance, and coverage region variances as well as image quality, limiting their robustness in real-world cross-modal image registration. In this paper, we propose a novel cross-modal descriptor learning method based on Equiangular Unit Basis Vectors (EUBVs), achieving robustness against significant radiometric, geometric, and temporal variations, as well as image noise.  


\subsection{Area-based Image Registration}
Area-based methods utilize entire images to estimate transformation parameters \cite{34}, differing from local feature-based methods where image feature extraction and similarity measurement are crucial. Hua et al. \cite{36} adapted mutual information to quantify similarity between different modalities. Xu et.al\cite{38} introduced a metric based on Jeffrey divergence to overcome limitations of mutual information based on KL divergence. Some methods measure the similarity in the frequency domain. Xue et.al\cite{39} proposed a frequency-based terrain feature alignment algorithm. To reduce data volume and computational costs in registration, Eldad et al. \cite{52} explored octrees. Self-similarity is also employed, for example, Hong et al. \cite{41} introduced a local self-similarity descriptor and shape similarity between objects, enhancing images resilient to radiation changes.



However, it is still an open problem to design a similarity metric for robust cross-modal registration. Moreover, area-based methods are often time-consuming and memory-intensive, especially for large-scale images. We propose a detection-free and memory-efficient grid-based registration framework which tiles large-scale images into patches and extracts their descriptors via the proposed network.

\subsection{End-to-End Deep Image Registration}
End-to-end methods for multimodal image registration differ from traditional feature-based and area-based approaches, which directly regress transformation parameters by training neural networks. Li et.al\cite{44} unified all matching tasks, including feature detection, feature descriptor, and parameter estimation, under an end-to-end framework through a multimodal image fusion network with a self-attention mechanism. Ye et.al\cite{43} devised a multi-scale image registration framework MU-Net based on unsupervised learning, comprising multiple stacked DNNs to predict registration parameters from coarse to fine levels. For end-to-end multimodal image registration, the choice of loss function is very important. Czolbe et.al\cite{74} proposed a new metric to measure the similarity of semantics, which is based on learned semantic features. 


However, current methods still struggle with large-scale image registration involving substantial geometric and radiometric differences, especially across airborne and spaceborne platforms. To address this challenge, we propose a new multimodal image registration framework, Grid-Reg, which is capable of handling images of arbitrary size and achieving accurate registration even under significant modality gaps and geometric differences.

\section{Methodology}
\begin{figure*}[!htbp]
\centering
\includegraphics[width=0.9\linewidth]{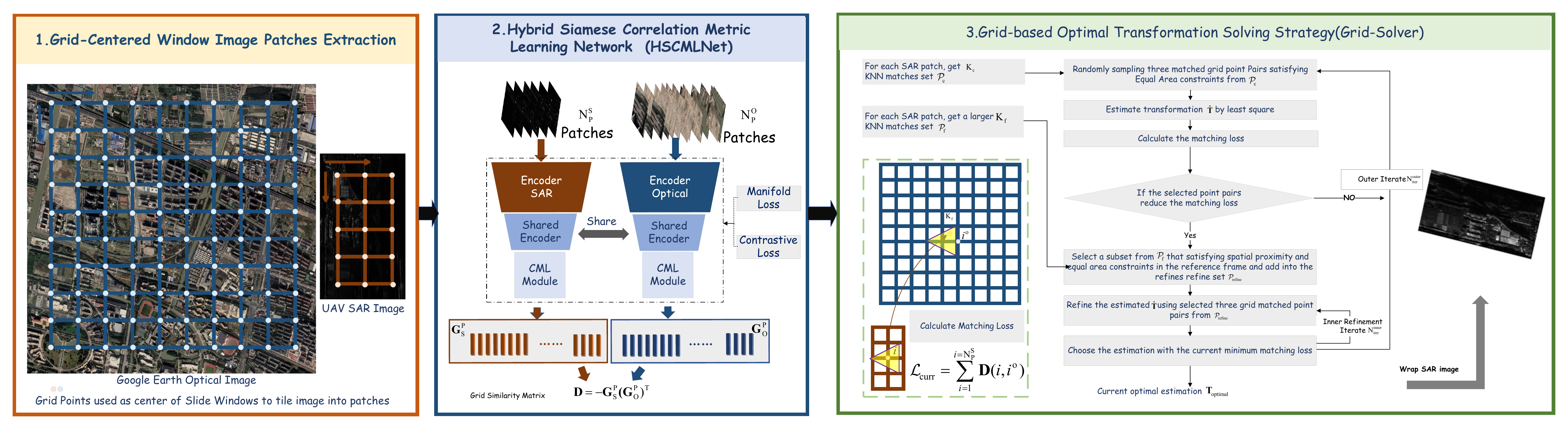}
\caption{Our Grid-Reg framework firstly parts the images into patches (1) and then extracts the patch descriptors by HSCMLNet(2). The Grid-Solver(3) searches for the optimal correspondences to achieve the minimal matching loss to estimate the registration transformation parameters.}
\label{overview_fig}
\end{figure*}


In Fig.~\ref{overview_fig}, our grid-based image registration framework is detector-free and comprises two main components: the Hybrid Siamese Correlation Metric Learning Network (HSCMLNet) for modality-invariant patch feature extraction and the grid-based transformation parameter solver (Grid-Solver). SAR and optical images are fistly tiled into patches {\small $\vec{\mathrm{P}}_{\mathrm{S}} \in \mathbb{R}^{\mathrm{N_P^S} \times \mathrm{H^P} \times \mathrm{W^P}}$} and {\small $\vec{\mathrm{P}}_{\mathrm{O}} \in \mathbb{R}^{\mathrm{N_P^O} \times \mathrm{H^P} \times \mathrm{W^P}}$} using a predefined sliding window {\small $\mathrm{H^P} \times \mathrm{W^P}$} with a fixed step size. Here, $N_P^{S}$ and $N_P^{O}$ are the total number of patches extracted from the SAR and optical images, respectively, and $H^{P}$, $W^{P}$ are the patch height and width. HSCMLNet extracts modality-invariant descriptors {\small $\vec{\mathrm{F}}_\mathrm{P}^\mathrm{S} \in \mathbb{R}^{\mathrm{N_P^S \times C}}$} and {\small $\vec{\mathrm{F}}_\mathrm{P}^\mathrm{O} \in \mathbb{R}^{\mathrm{N_P^O} \times \mathrm{C}}$} from SAR and optical patches, respectively, where $C$ is the descriptor dimension, and compute their similarity. The Grid-Solver then estimates optimal transform parameters {\small $\vec{\mathrm{T}}_{\mathrm{optimal}}$} by minimizing a global matching loss {\small $\mathcal{L}_{\mathrm{curr}}$} using a progressive dual-loop search strategy to identify optimal patch correspondences. In this strategy, the outer loop explores candidate grid point matches, while the inner loop refines the estimates by searching within an expanded candidate set guided by the outer-loop results. This dual-loop process mitigates the risk of local optima when relying solely on the outer loop. The procedure iterates until a predefined number of iterations is reached, and the final estimate is selected from the correspondences that yield the minimal global matching loss.
\subsection{Hybrid Siamese Correlation Metric Learning Network}\label{subsectionA}
In Fig.~\ref{overview_fig}, HSCMLNet includes a hybrid siamese module (HSModule) and a correlation metric learning module (CMLModule). Due to the different imaging mechanisms between SAR and optical sensors, SAR and optical patches exhibit significant differences in low-level features. Nevertheless, when covering the same geographic area, they inherently depict the same objects and thus share common high-level characteristics. To leverage this, we adopted a hybrid Siamese architecture (HSModule) that employs two modality-specific encoders at the lower layers to extract domain-dependent features, followed by a shared encoder to capture modality-invariant representations. In contrast to the traditional Siamese architecture, our HSModule mitigates feature conflicts arising from large cross-modal gaps, thereby enhancing robustness to image noise and temporal variations, and facilitating the extraction of reliable and discriminative matching features.The resulting feature maps are input through the CMLModule to generate final descriptors for SAR and optical patches. 


Traditional patch-level descriptor learning methods usually employ a contrastive-like loss, enforcing the similarity between positive patches to exceed that of the hardest negatives by a fixed margin. Although effective for classification tasks, these approaches often cause the learned descriptors to be overly constrained to specific local regions within a patch, thereby limiting their ability to capture the discriminative features that are essential for accurate matching. As a result, the similarity measurement becomes fragile, particularly when the patches are corrupted by noise or contain temporal changes. To address these issues, we introduce the Manifold-consistency EUBVs Reconstruction Loss (EUVBsLoss) to constrain local embeddings of the patches in a manifold space defined by equiangular unit basis vectors (EUBVs), promoting geometric and modality invariance. By doing so, our approach can enhance the robustness of learned patch-level descriptors against significant geometric and radiometric variations. 



\subsubsection{Hybrid Siamese Module(HSModule)}
\begin{figure}[!htbp]
\includegraphics[width=\linewidth]{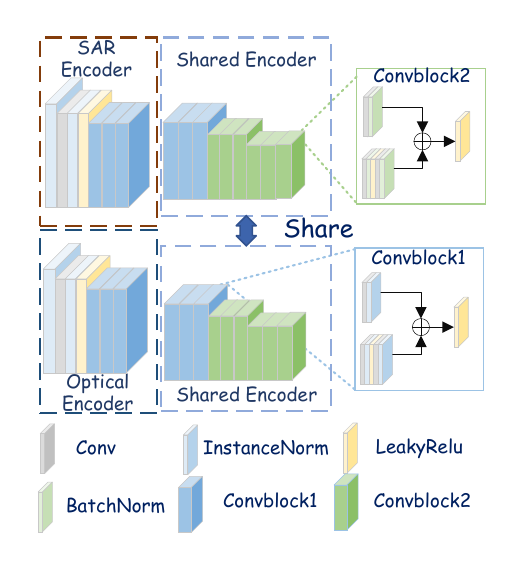}
\caption{The architecture of a hybrid Siamese module (HSModule). }
\label{HSNet}
\end{figure}
As shown in Fig.~\ref{HSNet}, the hybrid Siamese module combines two individual encoders and a shared encoder to extract features. The two individual encoders, each consisting of three Convblock1 units, extract low-level features from SAR and optical images, respectively. The resulting features from the SAR and optical branches are then passed to the shared encoder, which is composed of three Convblock1 and six Convblock2 units, both types being residual convolution blocks \cite{He2015DeepRL} with different normalization layers, producing feature maps {\small $\vec{\mathrm{F}}_{\mathrm{S}}, \vec{\mathrm{F}}_{\mathrm{O}} \in \mathbb{R}^{\mathrm{N_b} \times  C_F \times H_F \times W_F}$} corresponding to the SAR and optical patches, respectively, where $\mathrm{N_b}$ is the batch size.
\subsubsection{\textbf{Correlation Metric Learning Module(CMLModule)}}
\begin{figure*}[!htbp]
\includegraphics[width=\linewidth]{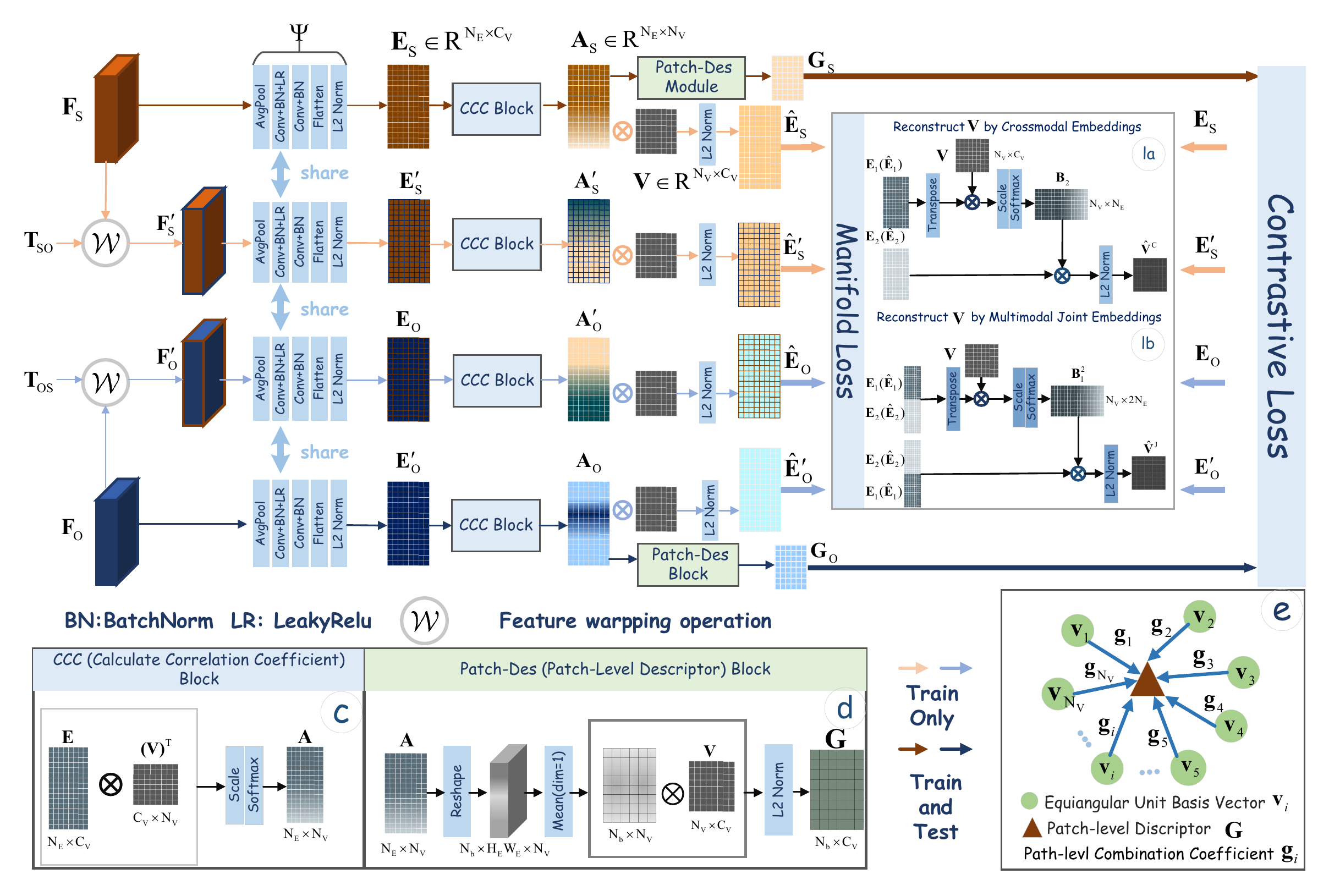}
\caption{The overview of Correlation Metric Learning Module based Local Manifold Equivalence Constraint(CMLModule);(1a) and (1b) introduce the calculation of the proposed manifold loss; (c): Calculate Correlation Coefficient(CCC) Block; (d):  Patch-Des(Patch-Level Descriptor) Block; (e): The generation of patch-level descriptor {\small $\vec{\mathrm{G}}$}.}
\label{metric_network}
\end{figure*}
To ensure robust patch-level descriptor and similarity computation, we design a correlation metric learning module (CMLModule), as illustrated in Fig.~\ref{metric_network}. While existing approaches primarily emphasize feature learning to separate positive patches from the most challenging negative ones, our CMLModule integrates the local embeddings of {\small $\vec{\mathrm{F}}_{\mathrm{S}}$} and {\small $\vec{\mathrm{F}}_{\mathrm{O}}$} into the holistic patch-level descriptors through an equiangular unit basis vectors (EUBVs) representation, which has been shown to improve performance for classification and metric learning compared to fully connected layers and classification heads\cite{7}.

Without loss of generality, given the output feature map of HSModule {\small $\vec{\mathrm{F}} \in \mathbb{R}^{\mathrm{N_b} \times  \mathrm{C_F} \times \mathrm{H_F} \times \mathrm{W_F}}$}(with the subscript defined contextually), it is further processed by the embedding layer $\Psi$ to obtain {\small $\vec{\mathrm{E}}=\Psi(\vec{F}) \in \mathbb{R}^{\mathrm{N}_E\times \mathrm{C}_E}$}, {\small $\mathrm{N}_E = \mathrm{N}_b \times \mathrm{H}_F\times \mathrm{W}_F$}. The embedding layer consists of an average pooling layer, a sequence of a Conv-BN-LR and a Conv–BN block, followed by a flattening operation and {\small $\mathrm{L_2}$} normalization. The flattening operation reshapes the feature map into a feature embedding matrix $\vec{E}$, after which {\small $\mathrm{L_2}$} normalization is applied along the feature dimension to ensure unit-length embeddings. 

In the following, we elaborate on the computation of patch descriptors using the EUBVs representation and the learning strategy for modality-invariant features for robust cross-modal matching. EUBVs are a set of unit basis vectors, denoted by {\small $\vec{\mathrm{V}} \in \mathbb{R}^{\mathrm{N_V \times C_V}}$},  which are pre-defined on a unit hypersphere of dimensions $\mathrm{C}_\mathrm{V}=\mathrm{C}_\mathrm{E}$ by 
\begin{equation}
\begin{aligned}
    &\vec{v}_i=\mathbf{V}_{i,:} = \frac{\mathbf{C}_{i,:}}{\|\mathbf{C}_{i,:}\|_2}, \quad i = 1, \dots, \mathrm{N_V},\\
    &\mathbf{C} = \mathbf{I}_{\mathrm{N_V}} - \frac{\mathbf{1}_{\mathrm{N_V \times N_V}}}{\mathrm{N_V}},
\end{aligned}
\end{equation}
where {\small \(\mathbf{I}_{\mathrm{N_V}} \in \mathbb{R}^{\mathrm{N_V \times N_V}}\)} denotes the {\small \(\mathrm{N_V}\)}-dimensional identity matrix, {\small \(\mathbf{1}_{\mathrm{N_V \times N_V}} \in \mathbb{R}^{\mathrm{N_V \times N_V}}\)} the all-ones matrix and {\small \(\|\cdot\|_2\)} the {\small $\mathrm{L_2}$} norm of a vector. The EUBVs {\small $\vec{\mathrm{V}}$} are designed to ensure the angle between any two vectors is equal to or close to 90 degree. For a local feature embedding at each position within a patch, we firstly 'project' it into the predefined manifold space by EUBVs and compute its correlation coefficients with the EUBV bases. This projection–correlation process can be implemented via the correlation coefficient calculation block (CCC) depicted in Fig.~\ref{metric_network}(c), formulated in matrix form as

\begin{equation}
\label{eq_A}
\vec{\mathrm{A}} = \mathrm{softmax}_{\mathrm{row}}\left(\frac{\vec{E}\vec{V}^{T}}{\tau_{A}}\right)
\end{equation}   
where $\mathrm{softmax}_{\mathrm{row}}(\cdot)$ denotes the row-wise softmax operation, the correlation coefficient matrix is denoted by {\small $\vec{A}\in \mathbb{R}^{\mathrm{N}_\mathrm{E}\times\mathrm{N}_\mathrm{V}}$},  {\small $\tau_{\mathrm{A}}$} represents the scale factor, with smaller values indicating greater attention to the most relevant EUBV. For each patch, the correlation coefficients across all spatial positions are subsequently aggregated to derive the patch-level descriptor $\vec{G}$ based on the EUBVs representation, as shown in Fig.~\ref{metric_network}(e) and computed as
\begin{equation}
    \vec{\mathrm{G}} = \sum_{i=1}^{\mathrm{N_v}} g_i\vec{\mathrm{v}}_i
\end{equation}
where the aggregation coefficients across all spatial positions {\small $\vec{g}=[g_i]_{i=1}^{\mathrm{N}_{\mathrm{V}}}$} are calculated by
\begin{equation}
     \vec{g} = \frac{1}{\mathrm{H_E}\mathrm{W_E}}\sum_{j=1}^{\mathrm{H_E}\mathrm{W_E}}\vec{\mathrm{A}}(\mathrm{patch\; index},j,:)
\end{equation}
where {\small $\vec{A}$} is reshaped as {\small $\mathrm{N_b}\times \mathrm{H_E}\mathrm{W_E}\times \mathrm{N_V}$} and is implemented by the Patch-Des block, as illustrated in Fig.~\ref{metric_network}(d).

\subsubsection{\textbf{Manifold-consistency EUBVs Reconstruction Loss}} 
Given the feature embeddings of the optical patches {\small $\vec{E}_\mathrm{O}=\Psi(\vec{F}_{\mathrm{O}})$}, SAR patches {\small $\vec{E}_\mathrm{S}=\Psi(\vec{F}_{\mathrm{S}})$}, and their corresponding registered embeddings {\small $\vec{E}_\mathrm{O}^{\prime}=\Psi(\mathcal{W}(\vec{T}_{\mathrm{OS}},\vec{F}_{\mathrm{O}}))$} and {\small $\vec{E}_\mathrm{S}^{\prime}=\Psi(\mathcal{W}(\vec{T}_{\mathrm{SO}},\vec{F}_{\mathrm{S}}))$}, to ensure that the desciptors remain robust against geometric and modality gap, it is desirable that 
\begin{align}
\label{e3}
      \vec{\mathrm{E}}_{\mathrm{S}} \approx \vec{\mathrm{E}}_{\mathrm{O}}^{\prime}, \;\; \vec{\mathrm{E}}_{\mathrm{O}} \approx \vec{\mathrm{E}}_{\mathrm{S}}^{\prime}
\end{align}
where {\small $\vec{\mathrm{T}}_{\mathrm{OS}}$} and {\small $\vec{\mathrm{T}}_{\mathrm{SO}}$} are the geometric transforms of registration, and {\small $\mathcal{W}(\cdot)$} is the wrapping operation. 

To achieve this cross-modal feature alignment, we project the registrated feature embedding pairs into a unified manifold space predefined by the EUBVs, where each EUBV can be regarded as a fixed cluster center. Within this unified space, the cluster centers derived from different modalities must be mutually aligned, and also consistent with the joint-modal cluster centers, which means that centers represented by the identical EUBVs can be reconstructed in a consistent manner. To realize this consistency, a necessary condition is that the representation coefficients computed from the embeddings of one modality for reconstructing the cluster centers can be directly transferred to the other modality to reconstruct the same cluster centers (i.e., the EUBVs). Furthermore, we enforce this consistency constraint within each mini-batch, thereby ensuring that the manifolds of both modalities are aligned almost everywhere. 

Without loss of generality, let {\small $\vec{E}_{1}\in\mathbb{R}^{\mathrm{N}_E\times C_E}$} and {\small $\vec{E}_{2}\in\mathbb{R}^{\mathrm{N}_E\times C_E}$} denote the embedding matrices of the registered patch pairs from the two modalities. From Equation~\ref{eq_A}, the embeddings {\small $\vec{E}_{1}$} and {\small $\vec{E}_{2}$} can be represented in terms of the EUBV bases as
\begin{equation}
\begin{aligned}
        \hat{\vec{\mathrm{E}}}_{1} = \mathrm{Norm}_{\mathrm{L_2, row}}\left(\vec{\mathrm{A}}_{1}\vec{\mathrm{V}}\right), & \;\; \hat{\vec{\mathrm{E}}}_{2} =\mathrm{Norm}_{\mathrm{L_2,row}} \left(\vec{\mathrm{A}}_{2}\vec{\mathrm{V}}\right)
\end{aligned}
\end{equation}
where $\mathrm{Norm}_{\mathrm{L_2, row}}(\cdot)$ denotes the row-wise {\small $\mathrm{L_2}$} normalization applied to each row vector. The cluster centers for each modality are derived by firstly computing the corresponding clustering coefficients as
\begin{equation}
\begin{aligned}
\vec{\mathrm{B}}_1 =\mathrm{softmax}_{\mathrm{row}}\left(\frac{\vec{\mathrm{V}}\vec{\mathrm{E}}_1^\mathrm{T}}{\tau_\mathrm{B}^{\mathrm{C}}}\right), &\;\;\vec{\mathrm{B}}_2 =\mathrm{softmax}_{\mathrm{row}}\left(\frac{\vec{\mathrm{V}}\vec{\mathrm{E}}_2^\mathrm{T}}{\tau_\mathrm{B}^{\mathrm{C}}}\right)
\end{aligned}
\end{equation}

\begin{figure}[!htbp]
\centering
\includegraphics[width=\linewidth]{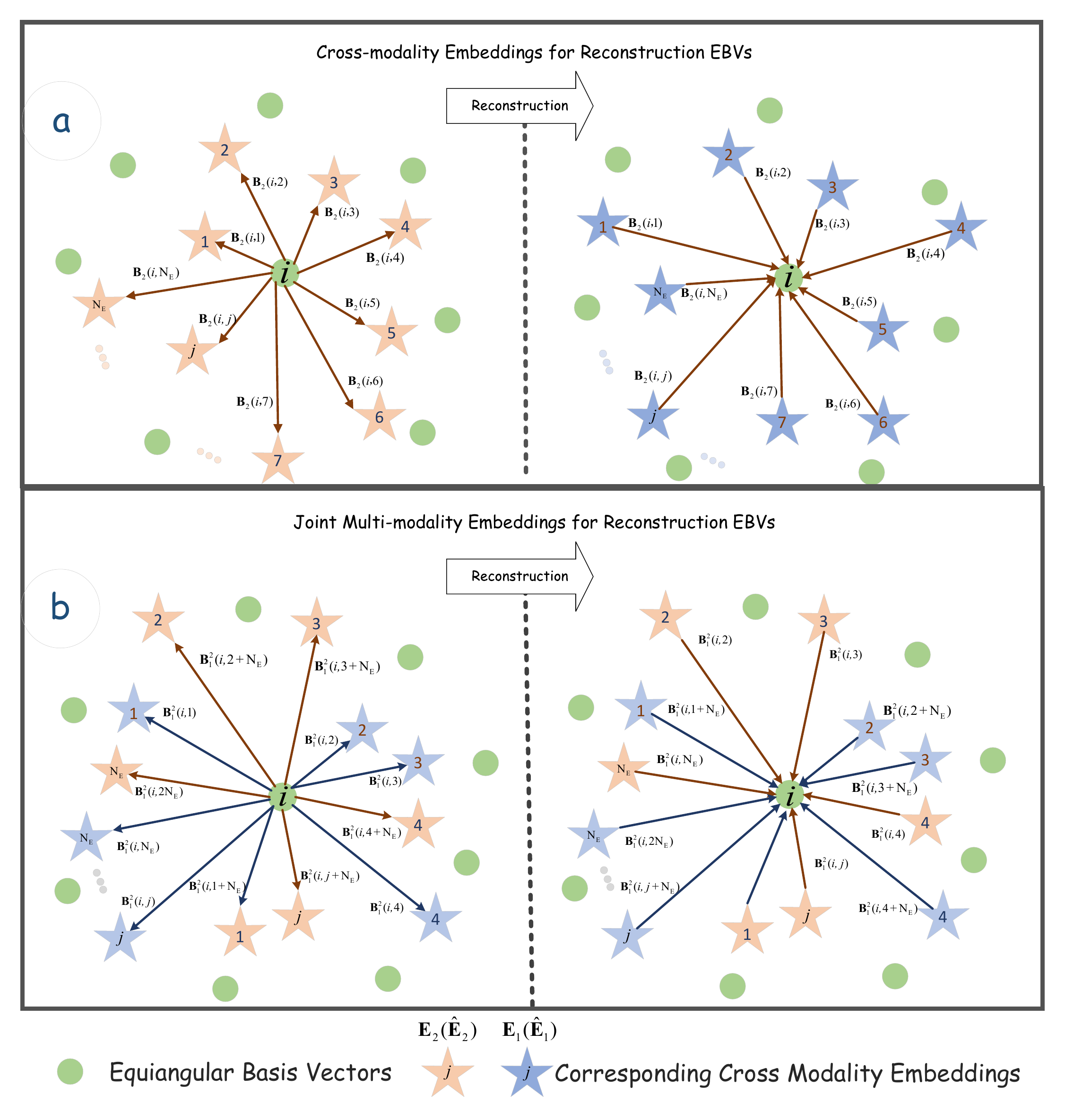}
\caption{EUBVs are reconstructed by local embeddings of one modality(a) with coefficients calculated from the the other modality and (b) the joint modalities.}
\label{LED1}
\end{figure}

Based on the bidirectional reconstruction mechanism, which aligns cross-modal embeddings within a shared EUBV-defined spherical manifold, as illustrated in Fig.~\ref{LED1}, the clustering coefficients derived from one modality (e.g., {\small $\vec{B}_2$}) are combined with the local embeddings from the other modality (e.g., {\small $\vec{E}_1$}) to reconstruct the same cluster centers (i.e.,the EUBVs), and vice versa, which can be formulated as 

\begin{equation}
\label{e_al}
\begin{split}
\hat{\vec{\mathrm{V}}}^{\mathrm{C}}(\vec{\mathrm{E}}_1,\vec{\mathrm{E}}_2) &= \mathrm{Norm}_{\mathrm{L_2, row}}(\vec{B}_2\vec{E}_{1}) \\
&= \mathrm{Norm}_{\mathrm{L_2, row}}\left(\mathrm{softmax}_{\mathrm{row}}\left(\frac{\vec{\mathrm{V}}\vec{\mathrm{E}}_2^\mathrm{T}}{\tau_\mathrm{B}^{\mathrm{C}}}\right)\vec{E}_{1}\right)
\end{split}
\end{equation}

This bidirectional reconstruction mechanism preserves not only the original embedding pairs {\small $\vec{E}_1, \vec{E}_2$}but also their newly induced representations {\small $\hat{\vec{E}}_1, \hat{\vec{E}}_2$} based on EUBVs, enhancing cross-modal alignment in both the original and transformed representational domains. Therefore, the cross-modality consistency EUBV reconstruction loss, EUBVsloss, aggregates all combinations of original and reconstructed embeddings across both modalities, ensuring that the cluster centers (EUBVs) remain consistent under bidirectional reconstruction. Given the cross-modal pairs {\small $\mathcal{S} = \{(\vec{\mathrm{E}}_{\mathrm{S}},\vec{\mathrm{E}}_{\mathrm{O}}^{\prime}),(\vec{\mathrm{E}}_{\mathrm{S}}^{\prime},\vec{\mathrm{E}}_{\mathrm{O}})\}$}, the cross-modal EUBVs reconstruction loss  is then formulated as

\begin{equation}
\begin{split}
\mathcal{L}_\mathrm{M}^{\mathrm{C}}
&= - \sum_{\scriptscriptstyle{(\vec{\mathrm{E}}_1,\vec{\mathrm{E}}_2) \in \mathcal{S}}} 
    \sum_{\scriptscriptstyle{ \tilde{\vec{\mathrm{E}}}_1 \in \{\vec{\mathrm{E}}_1, \hat{\vec{\mathrm{E}}}_1\}}}
\sum_{\scriptscriptstyle{ \tilde{\vec{\mathrm{E}}}_2 \in \{\vec{\mathrm{E}}_2, \hat{\vec{\mathrm{E}}}_2\}}}\\
&\quad\quad  \small{\mathrm{Tr}\left(\vec{\mathrm{V}}\!\left(\hat{\vec{\mathrm{V}}}^\mathrm{C}(\tilde{\vec{\mathrm{E}}}_1,\tilde{\vec{\mathrm{E}}}_2)\right)^{\mathrm{T}}\right)+\mathrm{Tr}\left(\vec{\mathrm{V}}\!\left(\hat{\vec{\mathrm{V}}}^\mathrm{C}(\tilde{\vec{\mathrm{E}}}_2,\tilde{\vec{\mathrm{E}}}_2)\right)^{\mathrm{T}}\right)}
\end{split}
\end{equation}

By projecting embeddings of both modalities into a common EUBV space, the proposed bidirectional reconstruction not only enables each modality to recover the EUBVs from the other, but also allows for joint reconstruction by combing embeddings and coefficients from both modalities, further reinforcing the alignment and consistency across modalities. In addition, owing to inherent modality gap, cross-modal reconstruction alone may yield suboptimal EUBV recovery. Therefore, we incorporate joint-modal EUBVs reconstruction loss, which further enhances the stability and robustness of the training process.


By concatenating {\small $\vec{\mathrm{E}}_1, \vec{E}_2$} into {\small $\vec{\mathrm{E}}_{1}^{2}=[\vec{E}_{1};\vec{E}_{2}]$}, and {\small $\vec{\mathrm{E}}_{2}^{1}=[\vec{E}_{2};\vec{E}_{1}]$}, the joint reconstruction coefficients can be computed as 
\begin{equation}
\vec{\mathrm{B}}_2^1 = \mathrm{softmax}_{\mathrm{row}}\left(\frac{\vec{\mathrm{V}}(\vec{\mathrm{E}}_2^1)^\mathrm{T}}{\tau_\mathrm{B}^\mathrm{J}}\right)
\end{equation}
 And as illustrated in Fig.~\ref{LED1}(b), the reconstructed EUBVs {\small $\hat{\vec{\mathrm{V}}}^\mathrm{J}(\vec{\mathrm{E}}_1,\vec{\mathrm{E}}_2)$} by joint embeddings of both modalies is computed as:
 \begin{equation}
\hat{\vec{\mathrm{V}}}^\mathrm{J}(\vec{\mathrm{E}}_1,\vec{\mathrm{E}}_2) = \mathrm{Norm}_{\mathrm{L2,row}}\left(\vec{\mathrm{B}}_2^{1}\vec{\mathrm{E}}_1^{2}\right) 
\end{equation} 
Thus, the joint multimodal consistency EUBVsLoss {\small $\mathcal{L}_\mathrm{M}^\mathrm{J}$} is then calculated as:
\begin{equation}
    \begin{aligned}
        \mathcal{L}_\mathrm{M}^\mathrm{J} = -\sum_{\scriptscriptstyle(\vec{\mathrm{E}}_1,\vec{\mathrm{E}}_2)\in S} \sum_{\scriptscriptstyle\tilde{\vec{E}}_{1}\in \{\vec{E}_1, \hat{\vec{E}}_1\}}\sum_{\scriptscriptstyle\tilde{\vec{E}}_{2}\in \{\vec{E}_2, \hat{\vec{E}}_2\}}& \small{\mathrm{Tr}\left(\vec{\mathrm{V}}(\hat{\vec{\mathrm{V}}}^\mathrm{J}(\tilde{\vec{{E}}}_1,\tilde{\vec{\mathrm{E}}}_2))^\mathrm{T}\right)}
    \end{aligned}
\end{equation}

\subsubsection{Contrastive Loss}
To enhance the discriminative capability of the patch-level descriptors, we employ a contrastive loss that minimizes the distance between positive sample pairs while maximizing that between negative pairs in the feature space. The contrastive loss defined over {\small $\vec{\mathrm{G}}{\mathrm{S}}$} and {\small $\vec{\mathrm{G}}{\mathrm{O}}$} is given by:
\begin{equation}
\mathcal{L}_{\mathrm{C}} = \sum_{i=1}^{i=\mathrm{N}_b}-\ln(\frac{\exp(\vec{\mathrm{G}}_{\mathrm{S}}^{i}(\vec{\mathrm{G}}_{\mathrm{O}}^{i})^\mathrm{T})/\tau_\mathrm{C})}{\mathrm{Z}})
\end{equation}
where {\small $\mathrm{Z}$} is the normalization term, calcuated as  
\begin{align}
\begin{aligned}
\mathrm{Z} &= \exp(\vec{\mathrm{G}}_{\mathrm{S}}^{i}(\vec{\mathrm{G}}_{\mathrm{O}}^{i})^\mathrm{T})/\tau_\mathrm{C})\\
&+\sum_{j \ne i} \exp(\vec{\mathrm{G}}_{\mathrm{S}}^{j}(\vec{\mathrm{G}}_{\mathrm{O}}^{i})^\mathrm{T}+\alpha)/\tau_\mathrm{C})+ \exp(\vec{\mathrm{G}}_{\mathrm{O}}^{j}(\vec{\mathrm{G}}_{\mathrm{O}}^{i})^\mathrm{T}+\alpha)/\tau_\mathrm{C})\\
&+\exp(\vec{\mathrm{G}}_{\mathrm{S}}^{j}(\vec{\mathrm{G}}_{\mathrm{S}}^{i})^\mathrm{T}+\alpha)/\tau_\mathrm{C})+ \exp(\vec{\mathrm{G}}_{\mathrm{O}}^{i}(\vec{\mathrm{G}}_{\mathrm{S}}^{j})^\mathrm{T}+\alpha)/\tau_\mathrm{C})
\end{aligned}
\end{align}  

Finally, the overall training loss {\small $\mathcal{L}$} in our proposed HSCMLNet is then defined as the sum of the {\small $\mathcal{L}_\mathcal{M}^\mathrm{C}$}, {\small $\mathcal{L}_\mathcal{M}^\mathrm{J}$}, and {\small $\mathcal{L}_\mathrm{C}$} given by:
\begin{equation}
\mathcal{L} = \mathcal{L}_\mathrm{M}^\mathrm{C}+\mathcal{L}_\mathrm{M}^\mathrm{J}+\mathcal{L}_\mathrm{C}
\end{equation} 

\subsection{Grid-based Optimal Transformation Solving Strategy} \label{subsectionB}
Due to the significant radiometric and geometric differences between large-scale UAV SAR and optical images across platforms, feature-based methods often struggle with reliable keypoints matching, resulting in a high number of outlier correspondences. Based on the aforementioned cross-modal patch-level descriptor extraction and similarity computation,  we introduce a grid-based patch matching strategy to identify optimal grid-point correspondences for estimating registration parameters, avoiding fragile keypoint detection and matching across modalities. Given patch-level descriptors of SAR and Optical image output from the proposed HMSCMLNet, {\small $\vec{G}^{\mathrm{P}}_{\mathrm{S}}\in \mathbb{R}^{\mathrm{N_S^P\times C_V}}$} and {\small $\vec{G}^{\mathrm{P}}_{\mathrm{O}}\in \mathbb{R}^{\mathrm{N_O^P\times C_V}}$}, the global matching objective is defined as 
\begin{equation}
\label{eqn:similarity_D}
\begin{split}
\vec{\mathrm{D}} = -\frac{\vec{\mathrm{G}}_\mathrm{S}^\mathrm{P}{(\vec{\mathrm{G}}_\mathrm{O}^\mathrm{P})}^\mathrm{T}}{\|\vec{\mathrm{G}}_\mathrm{S}^\mathrm{P}{(\vec{\mathrm{G}}_\mathrm{O}^\mathrm{P})}^\mathrm{T}\|_2}, &\;\; \vec{\mathrm{D}} \in [-1, 1]^{\mathrm{N_S^P} \times \mathrm{N_O^P}}
\end{split}
\end{equation}
where {\small $\mathrm{N_S^P}$, $\mathrm{N_O^P}$} are the number of tiled patches from the input SAR and optical image, respectively.

To identify the optimal patch correspondences, we introduce a dual-loop search strategy that minimizes the matching objective, as illustrated in Fig.~\ref{got} 



\subsubsection{Gridizing Image}
As shown in Fig.~\ref{overview_fig}(A), the input SAR of size {\small $\mathrm{H^{S}}\times\mathrm{W^{S}}$} and optical image of size {\small $\mathrm{H^{O}}\times\mathrm{W^{O}}$} are firstly partitioned into patches into patches of size {\small $\mathrm{H}^{\mathrm{P}}\times \mathrm{W}^{\mathrm{P}}$} using a sliding window with a $\mathrm{Step}$. Each patch center serves as a grid point associated with the corresponding patch-level descriptor. In our curated dataset where SAR and optical images have a resolution of about  $0.51m$, we set {\small $\mathrm{H^P = W^P = 256}$}, ensuring each each patch adequately covers a medium-sized building.


\subsubsection{Obtaining the Initial Set of Matched Grid-Point Pairs}
After cropping, the generated patches are fed into the trained HSCMLNet to obtain patch-level descriptors and compute their similarity, yielding the distance matrix {\small \( \vec{\mathrm{D}} \)}.


In the following, for each grid point {\small $(x_i^{\mathrm{s}},y_i^{\mathrm{s}})$} in SAR image, we identify its {\small $\mathrm{K_c}$} and {\small $\mathrm{K_f}$} nearest neighbor grid points in optical images based on their similarity, forming corresponding grid point set {\small ${\mathcal{P}}_\mathrm{c}$} and {\small ${\mathcal{P}}_\mathrm{f}$}, respectively.

As the coverage of the reference image increases, similar structures become more frequent. In this paper, we set {\small $\mathrm{K_c}$} and {\small $\mathrm{K_f}$} to be proportional to the SAR and optical coverage area ratios, given by
\begin{equation}
\begin{split}
\mathrm{K_c} = \mathrm{K}\sqrt{\frac{\mathrm{H^O}\times \mathrm{W^O}}{\mathrm{H^S}\times \mathrm{W^S}}}\frac{\mathrm{Step}}{16},&\;\;
\mathrm{K_f} = 4\mathrm{K_c}
\end{split}
\end{equation}

In the subsequent dual-loop search stage, {\small ${\mathcal{P}}_\mathrm{c}$} is employed in the outer loop to estimate transformation parameters coarsely, while {\small ${\mathcal{P}}_\mathrm{f}$} are then utilized in inner-loop to refine the outer-loop estimate. Assigning {\small ${\mathcal{P}}\mathrm{c}$} a smaller cardinality than {\small ${\mathcal{P}}\mathrm{f}$} facilitates a faster search toward a reliable initial estimate, while the larger {\small ${\mathcal{P}}_\mathrm{c}$} provides denser and more accurate local correspondences for refinement, simultaneously mitigating the risk of convergence to suboptimal solutions when relying solely on {\small ${\mathcal{P}}\mathrm{c}$}.

\begin{figure*}[!htbp]
\includegraphics[width=\linewidth]{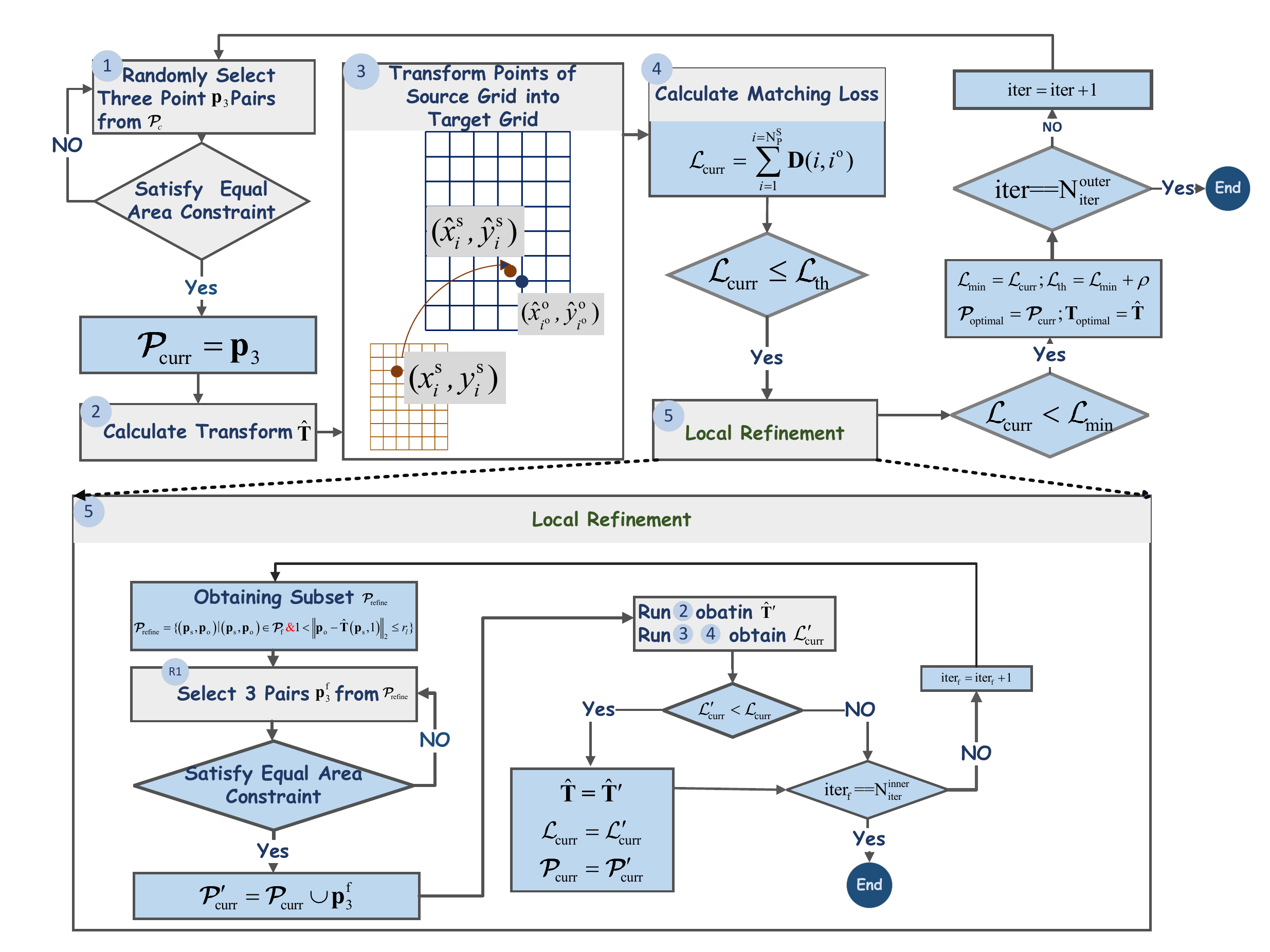}
\caption{{The flowchart of dual-loop search strategy to identify the optimal patch correspondences achieving the minimal matching objective.}}
\label{got}
\end{figure*}

\subsubsection{\textbf{Estimating Global Optimal Transformation by progressive Dual-Loop Search}}
As shown in Fig.~\ref{got}, we iteratively identify the optimal correspondences to minimize the grid matching objective {\small $\mathcal{L}_{\mathrm{min}}$} to find the optimal transformation {\small $\vec{\mathrm{T}}_{\text{optimal}}$}. It involves an outer-loop search and an inner-loop search. In the outer-loop search, the three point pairs {\small $\vec{p}_3$} that satisfy the triangle area constraint are chosen from {\small $\vec{\mathcal{P}}_\mathrm{c}$} to estimate the coarse transform matrix {\small $\hat{\vec{\mathrm{T}}}$}. The triangle-area constraint requires that the ratio of the areas of the triangles formed by {\small $\vec{p}_3$} in the SAR and optical images falls within specified lower and upper thresholds to discard abnormal correspondences. We then search for the optimal $\vec{p}_3$ in {\small $\vec{\mathcal{P}}_\mathrm{c}$} by evaluating the correspondences associated with the estimated  {\small $\hat{\vec{\mathrm{T}}}$} based on the proposed matching objective, i.e., Equation.~(\ref{eqn:similarity_D}).
For each grid point {\small $(x_i^{\mathrm{s}}, y_i^\mathrm{s})$} in the SAR image  grid, the index {\small $i^{\mathrm{o}}$} of the corresponding grid point {\small $(x_{i^{\mathrm{o}}}^\mathrm{o}, y_{i^\mathrm{o}}^\mathrm{o})$} in the referenced optical image grid is calculated as follows:
\begin{equation}
\begin{array}{l}
(\hat{x}_i^{\mathrm{o}},\hat{y}_i^\mathrm{o})=\hat{\vec{\mathrm{T}}}(x_i^\mathrm{s},y_i^\mathrm{s},1.0)\\
i_x^\mathrm{o} = \text{clip}(\text{round}((\hat{x}_i^\mathrm{o}-\mathrm{W^P}//2)/\mathrm{Step}),0,\mathrm{N_W^O}-1)\\
i_y^\mathrm{o} = \text{clip}(\text{round}((\hat{y}_i^\mathrm{o}-\mathrm{H^P}//2)/\mathrm{Step}),0,\mathrm{N_H^O}-1)\\
i^\mathrm{o} = i_y^\mathrm{o} \times \mathrm{N_W^O} + i_x^\mathrm{o}
\end{array}
\end{equation}
The evaluation objective of matching using the current estimation parameters is then defined as:
\begin{equation}
\label{e8}
\mathcal{L}_{\mathrm{curr}} = \sum_{i=1}^{i=\mathrm{N_S^P}}\vec{\mathrm{D}}(i,i^\mathrm{o})
\end{equation}

If {\small $\mathcal{L}_{{\mathrm{curr}}}$} is not greater than the threshold {\small $\mathcal{L}_{\mathrm{th}}$}, the inner-loop is conducted to refine the current outer-loop estimate. At each iteration in the inner-loop search, a refined subset of point pairs, {\small ${\mathcal{P}}_{\mathrm{refine}}$} is selected from  {\small ${\mathcal{P}}_{\mathrm{f}}$} which are satisfying 
\begin{equation}
\scriptstyle 
{\mathcal{P}}_{\mathrm{refine}}= \{(\vec{\mathrm{p}}_\mathrm{s},\vec{\mathrm{p}}_\mathrm{o})|(\vec{\mathrm{p}}_\mathrm{s},\vec{\mathrm{p}}_\mathrm{o}) \in {\mathcal{P}}_{\mathrm{f}} \, \& \, 1<\|\vec{\mathrm{p}}_\mathrm{o}-\hat{\vec{\mathrm{T}}}(\vec{\mathrm{p}}_\mathrm{s},1.0)\|_2 \leq \mathrm{r_l}\}
\end{equation}
where {\small $\vec{\mathrm{p}}_\mathrm{s}$} is the grid point of the SAR image, and {\small $\vec{\mathrm{p}}_\mathrm{o}$} is its corresponding grid point of the optical image. Then, we add three matched point pairs {\small $\vec{\mathrm{p}}_3^\mathrm{f}$} from {\small ${\mathcal{P}}_{\mathrm{refine}}$} satisfying the area constraint to {\small ${\mathcal{P}}_{\mathrm{curr}}$} to estimate the transformation {\small $\hat{\vec{\mathrm{T}}}^{\prime}$}, and evaluate the estimate using {\small $\mathcal{L}_{{\mathrm{curr}}}$}.

The outer loop and inner loop are executed until a predefined number of iterations is reached. The correspondence set associated with the transformation estimate that achieves the minimum global loss {\small $\mathcal{L}_{\mathrm{curr}}$} is then selected as the optimal estimate. Additionally, a final refinement step can be performed in a manner similar to the local refinement. The main flowchart of the grid-solver algorithm is illustrated in Fig.~\ref{got}.

\section{Experiments and Results}
\subsection{Experiment Dataset}
Due to the lack of publicly available challenging large-scale SAR and optical image registration datasets across platforms, we constructed a benchmark dataset with Drone-based MiniSAR and optical images from Google Earth. Our dataset includes 29 image subsets, each containing SAR images and multiple optical images captured at different times from Google Earth. Compared with the existing datasets, our dataset exhibits significant geometric and modal differences between SAR and optical images as well as temporal changes. Moreover, the optical reference images usually cover a larger geographic area than the corresponding SAR images, containing similar structures, particularly in densely urban regions. To evaluate the impact of the overlap area ratios on registration performance, we categorized the test samples into four difficulty levels, as detailed in Table \ref{tab_test_level}. Here, RSO denotes the ratio of the SAR image coverage area to that of the optical reference image, while ROA refers to the ratio of the overlapping area between SAR and optical images to the total SAR image coverage area. To access registration robustness,  each SAR image in the test set is evaluated with multiple optical images acquired under diverse conditions (referred to Table \ref{tab_registration_data}).

In our dataset, the ground-truth affine transformations are computed by manually annotating control points for each image pair. 17 image subsets, including diverse scenes like campuses, suburbs, highways, and urban areas, are chosen for training and the remaining for testing, ensuring no overlap between the training and testing sets. 

\begin{table}[!htbp]
	\caption{Four Difficulty Levels in Our Test Dataset}
	\centering
	\begin{tabular}{c|c|c|c|c}
		\hline
		Level  &$\mathrm{L}_{-1}$  &$\mathrm{L}_0$ &$\mathrm{L}_1$ &$\mathrm{L}_2$\\
		\hline
		ROA &\makecell{[0.6,0.8]} &\makecell{1.0}&\makecell{1.0}&\makecell{1.0}\\
		\hline
		RSO &\makecell{4.0} &\makecell{4.0}&\makecell{9.0}&\makecell{14.0}\\
		\hline
	\end{tabular}
    \label{tab_test_level}
\end{table}

\begin{table}[!htbp]
	\caption{The information of our test dataset for large-scale multimodal image registration task(the number of pairs $n\times4$, where $n$ represents there are $n$ pairs in test ID $i$ at each difficulty level, and $4$ is the number of difficulty levels.)}
	\centering
	\begin{tabular}{c|c|c|c}
		\hline
		ID & SAR Size &\makecell{The Size of Each Optical\\($L{-1}$,$\mathrm{L}_\mathrm{0}$,$\mathrm{L}_\mathrm{1}$,$\mathrm{L}_\mathrm{2}$))} & \makecell{Number of Pairs\\ each difficulty level} \\
		\hline
		0 &\makecell{906$\times$891} &\makecell{$1797^2$,$1797^2$,$2696^2$,$3362^2$} &\makecell{21$\times$4}\\
		\hline
		1 &\makecell{1160$\times$1309} &\makecell{$2465^2$,$2465^2$,$3697^2$,$4611^2$} &\makecell{10$\times$4}\\
		\hline
		2 &\makecell{1001$\times$1190} &\makecell{$2183^2$,$2183^2$,$3275^2$,$4084^2$} &\makecell{10$\times$4}\\
		\hline
		
		3 &\makecell{1881$\times$450} &\makecell{$1841^2$,$1841^2$,$2761^2$,$3443^2$} &\makecell{10$\times$4}\\
		\hline
		
   		4 &\makecell{1288$\times$1220} &\makecell{$2508^2$,$2508^2$,$3761^2$,$4580^2$} &\makecell{10$\times$4}\\
		\hline
		
		5 &\makecell{975$\times$1659} &\makecell{$2544^2$,$2544^2$,$3816^2$,$4759^2$} &\makecell{10$\times$4}\\
		\hline
		
		6 &\makecell{723$\times$1985} &\makecell{$2396^2$,$2396^2$,$3594^2$,$4483^2$} &\makecell{10$\times$4}\\
		\hline
		7 &\makecell{1881$\times$503} &\makecell{$1946^2$,$1946^2$,$2919^2$,$3640^2$} &\makecell{11$\times$4}\\
		\hline
		8 &\makecell{707$\times$1264} &\makecell{$1891^2$,$1891^2$,$2836^2$,$3538^2$} &\makecell{10$\times$4}\\
		\hline
		9 &\makecell{712$\times$1948} &\makecell{$2356^2$,$2356^2$,$3534^2$,$4407^2$} &\makecell{20$\times$4}\\
		\hline
		10 &\makecell{460$\times$1755} &\makecell{$1797^2$,$1797^2$,$2696^2$,$3362^2$} &\makecell{11$\times$4}\\
		\hline
		11 &\makecell{728$\times$1828} &\makecell{$2308^2$,$2308^2$,$3461^2$,$4317^2$} &\makecell{21$\times$4}\\
		\hline
	\end{tabular}\label{tab_registration_data}
\end{table}

\subsection{Baseline Methods}

On one hand, we compare our HSCMLNet method with the following leading descriptor learning methods to demonstrate its effectiveness:
\begin{itemize}
\item HardNet \cite{10} emphasizes the distance between positive and hardest negative samples, using a dual sample approach \cite{47} to prevent over-fitting.
\item SosNet \cite{11} employs First Order Similarity Loss and Second Order Similarity Regularization (SOSR).
\item HyNet \cite{9} enhances local patch matching by focusing on relative distances among descriptors.
\item SDGMNet \cite{49} improves triplet loss with dynamic gradient modulation techniques.
\end{itemize}
These networks utilize L2Net as the backbone, which is originally designed for small grayscale image patches. To adapt L2Net for input patches of size {\small $\mathrm{256} \times \mathrm{256}$}, we modify its architecture accordingly. Since the input of L2Net is a grayscale image, we convert the RGB optical images into grayscale and employ a Siamese architecture to extract features from SAR and optical patches.
To evaluate the effectiveness of the proposed Grid-Solver, we compare it with a feature-based parameter estimation method (Feat-Solver). In Feat-Solver, keypoints are detected using ORB’s keypoint detection method, while the descriptors for these keypoints are obtained from the networks under comparison. These descriptors are then matched to estimate affine transformation parameters via RANSAC. For a fair comparison, the area constraint is applied after the random sampling step and before the model estimation step of RANSAC. If the constraint is violated, model estimation is skipped, effectively filtering out abnormal matches.
\begin{figure*}[!htbp]
	\centering
	\includegraphics[width=0.9\linewidth]{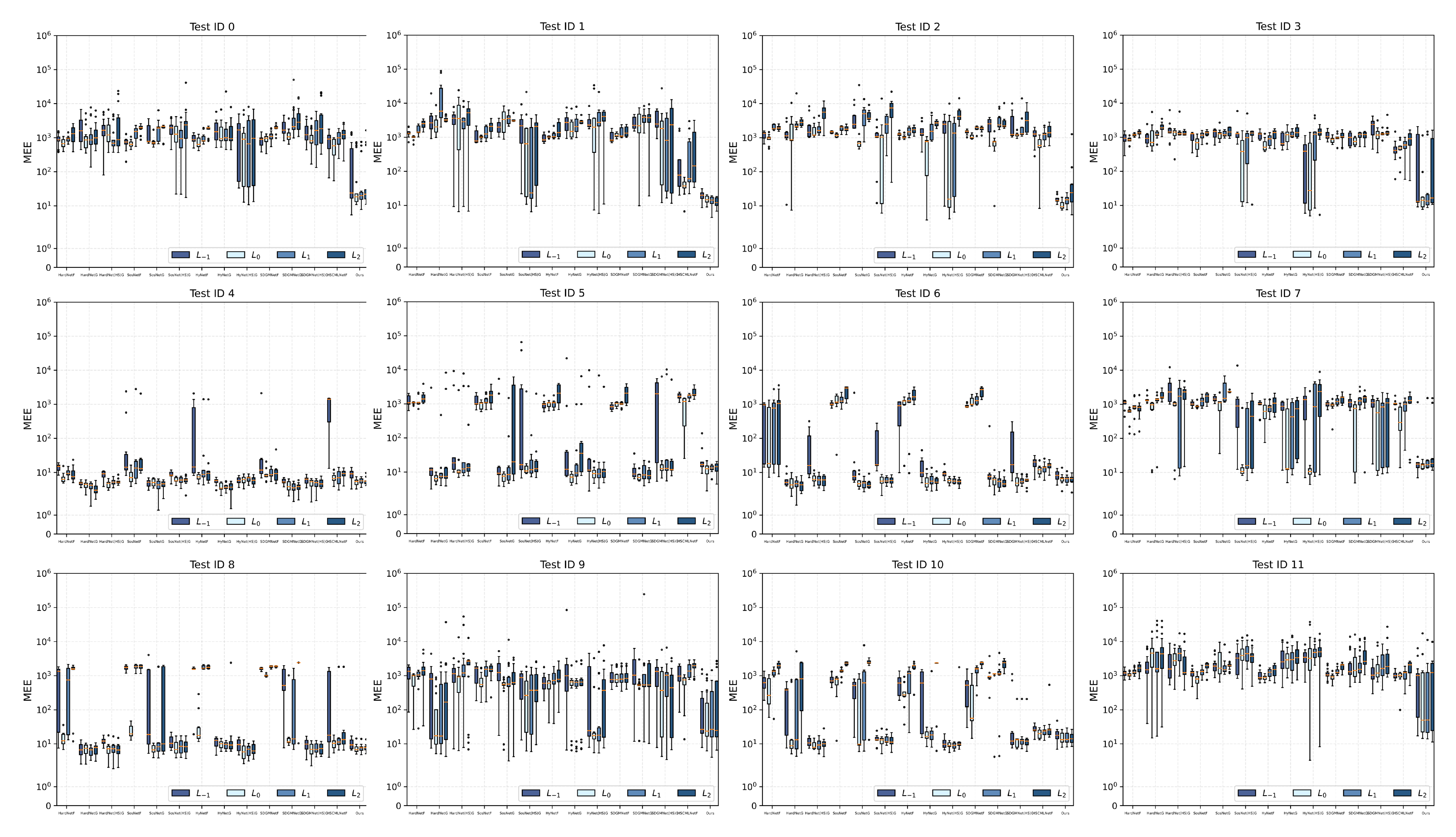}
	\centering
	\caption{The MEE of different methods on four levels for each test, which include  HardNet, SosNet, HyNet, SDGMNet, and HSCMLNet with  Feat-Solver(F) and Grid-Solver(G).}\label{box_compare}
\end{figure*}
\subsection{Implement Details}
\subsubsection{Training}
We adopt Adam with a learning rate {\small $2.5e^{-4}$} as the optimizer to train the networks. The batch size is set to 512. We apply data augmentation of random rotation and flip for  SAR and optical patches. All networks are trained using the same training parameters and data augmentation.
\subsubsection{Testing}
The number of iterations is a crucial factor influencing the performance of the algorithm. As the coverage area of the reference image expands, the difficulty of registration increases. Therefore, we set the number of iterations as {\small $\mathrm{N}_{\mathrm{iter}}^{\mathrm{outer}} = \beta \times \frac{\mathrm{H^O}\times \mathrm{W^O}+\mathrm{H^S\times W^S}}{2.0}$}, where {\small $\beta$} is used to adjust {\small $\mathrm{N}_{\mathrm{iter}}^{\mathrm{outer}}$} to balance registration accuracy and efficiency, and set {\small $\beta=1.0$} and {\small $\beta=2.0$} for Grid-Solver and Feat-Solver, respectively. For the Grid-Solver, we set {\small $\mathrm{Step}$} to 16. At the inner-loop refinement stage, we set the {\small $r_\mathrm{l}$}  as {\small $\min\{\mathrm{4} \times \mathrm{Step},\mathrm{100}\}$}.
\subsection{Evaluation Metrics}
We utilize the success rate\cite{67,24,69} to evaluate different methods, which is defined as
\begin{equation}
\mathrm{Reg}_\mathrm{sr}^\mathrm{th} = \frac{\mathrm{N}_{\mathrm{success}}}{\mathrm{N}_{\mathrm{images}}}
\end{equation}
where {\small $\mathrm{N}_{\mathrm{images}}$} is the total number of image pairs in the testing set, and {\small $\mathrm{N}_{\mathrm{success}}$} is the number of the successfully registered image pairs whose median error (MEE) are less than or equal to {\small $\mathrm{th}$} pixels. Since different tasks require different registration accuracy standards, we evaluate the registration success rate under four threshold settings: {\small $\mathrm{th} \in \{25, 50, 75, 100\}$} pixels, enabling a fair comparison across methods. The MEE is calculated as:
\begin{equation}
\small
    \label{rmse}
    \mathrm{MEE} = \mathrm{median} \{ || \hat{\vec{\mathrm{T}}}(x_i^s, y_i^s,1.0) - \vec{\mathrm{T}}(x_i^s, y_i^s,1.0) )||_2 \}_{i=1}^{\mathrm{N_p^k}}
\end{equation}
where {\small $(x_i^s,y_i^s) \in \{(x_i^s,y_i^s)|((x_i^s,y_i^s)\in \mathrm{I^s} \& T(x_i^s,y_i^s)\in \mathrm{I^t})\}$}
,{\small $\mathrm{I^s}$} denotes the source image, {\small $\mathrm{I^t}$} the reference image, {$\mathrm{N_p^k}$} the number of pixels in SAR image of the {$\mathrm{k}^{\mathrm{th}}$} test pair, {\small $\vec{\mathrm{T}}$} and {\small $\hat{\vec{\mathrm{T}}}$} are the ground-truth and estimated transform parameters, respectively.
\subsection{Comparison of Image Registration results}
\begin{figure}[!htbp]
	\centering
	\includegraphics[width=0.9\linewidth,height=50mm]{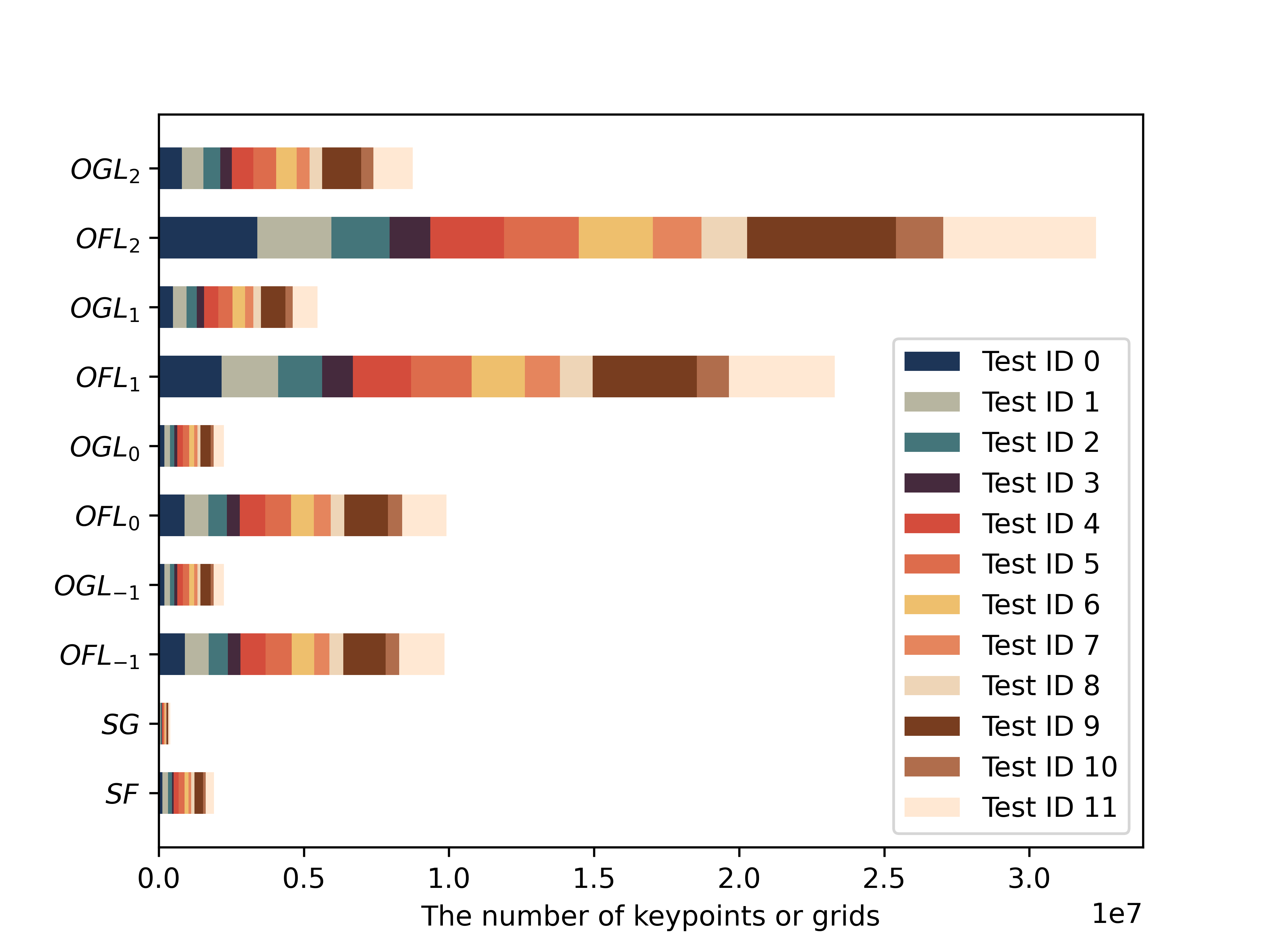}
	\centering
	\caption{Comparison of keypoints or grids between SAR and optical images at four levels ("O" for optical images, "S" for SAR images, "F" for Feat-Solver, "G" for Grid-Solver).) }\label{keypoints_compare}
\end{figure}

\begin{table}[!htbp]
 	\caption{Comparing Our Results with Feature-based methods on Four Levels. "F": Feat-Solver, "G": Grid-Solver.}
	\centering
 	\begin{tabular}{c|c|c|c|c|c}
 		\hline
 		\multicolumn{2}{c|}{Method} &\multicolumn{4}{c}{$\mathrm{L}_\mathrm{-1}$}  \\
 		\hline
 		\multicolumn{2}{c|}{} &{\small $\mathrm{Reg_{sr}^{25}}$} &{\small $\mathrm{Reg_{sr}^{50}}$} &{\small $\mathrm{Reg_{sr}^{50}}$} &{\small $\mathrm{Reg_{sr}^{100}}$}\\
 		\hline
            \multicolumn{2}{c|}{HardNetF}   &\makecell{11.04\%} &\makecell{11.69\%}&\makecell{11.69\%}&\makecell{12.34\%}\\
            \hline
     	\multicolumn{2}{c|}{SosNetF}  &\makecell{4.55\%} &\makecell{5.19\%} &\makecell{5.19\%} 
            &\makecell{5.19\%} \\
     	\hline
		  \multicolumn{2}{c|}{HyNetF}&\makecell{5.84\%} &\makecell{6.49\%} &\makecell{7.14\%} 
            &\makecell{7.14\%} \\
 		\hline
 		\multicolumn{2}{c|}{SDGMNetF}&\makecell{4.54\%} &\makecell{7.79\%} 
            &\makecell{7.79\%}&\makecell{7.79\%}\\
 		\hline
 		\multicolumn{2}{c|}{Ours}  &\makecell{\textbf{65.58\%}} 
            &\makecell{\textbf{75.32\%}}&\makecell{\textbf{76.62\%}}&\makecell{\textbf{77.92\%}}\\
 		\hline
 		\multicolumn{2}{c|}{} &\multicolumn{4}{c}{{\small $\mathrm{L_0}$}}  \\
 		\hline
 		\multicolumn{2}{c|}{} &{\small $\mathrm{Reg_{sr}^{25}}$} &{\small $\mathrm{Reg_{sr}^{50}}$} &{\small $\mathrm{Reg_{sr}^{75}}$} &{\small $\mathrm{Reg_{sr}^{100}}$}\\
 		\hline
            \multicolumn{2}{c|}{HardNetF}&\makecell{16.88\%} &\makecell{18.18\%} &\makecell{19.48\%}&\makecell{20.13\%}\\
            \hline
 		\multicolumn{2}{c|}{SosNetF}&\makecell{11.69\%} &\makecell{14.29\%} &\makecell{14.29\%}  &\makecell{14.29\%}\\
 		\hline
 		\multicolumn{2}{c|}{HyNetF} &\makecell{11.69\%} &\makecell{13.64\%} &\makecell{13.64\%} &\makecell{14.29\%}\\
 		\hline
 		\multicolumn{2}{c|}{SDGMNetF}&\makecell{7.14\%} &\makecell{8.44\%} &\makecell{10.39\%}&\makecell{10.39\%}\\
 		\hline
 		\multicolumn{2}{c|}{Ours}  &\makecell{\textbf{80.52\%}} &\makecell{\textbf{85.71\%}}&\makecell{\textbf{85.71\%}}&\makecell{\textbf{87.01\%}}\\
 		\hline
 		\multicolumn{2}{c|}{} &\multicolumn{4}{c}{{\small $\mathrm{L_1}$}}  \\
 		\hline
 		\multicolumn{2}{c|}{} &{\small $\mathrm{Reg_{sr}^{25}}$} &{\small $\mathrm{Reg_{sr}^{50}}$} &{\small $\mathrm{Reg_{sr}^{75}}$} &{\small $\mathrm{Reg_{sr}^{100}}$}\\
 		\hline
    	\multicolumn{2}{c|}{HardNetF}&\makecell{11.69\%} &\makecell{12.99\%} &\makecell{13.64\%}&\makecell{13.64\% }\\
    	\hline
 		\multicolumn{2}{c|}{SosNetF} &\makecell{5.84\%} &\makecell{5.84\%} &\makecell{5.84\%} &\makecell{5.84\% }\\
 		\hline
 		\multicolumn{2}{c|}{HyNetF} &\makecell{6.49\%} &\makecell{8.44\%} &\makecell{8.44\% }&\makecell{8.44\%}\\
 		\hline
 		\multicolumn{2}{c|}{SDGMNetF} &\makecell{6.49\%} &\makecell{6.49\%} &\makecell{6.49\%} &\makecell{6.49\% }\\
 		\hline
 		\multicolumn{2}{c|}{Ours}  &\makecell{\textbf{78.57\%}} &\makecell{\textbf{83.12\%}}&\makecell{\textbf{84.42\%}}&\makecell{\textbf{84.42\%}}\\
 		\hline
		 \multicolumn{2}{c|}{} &\multicolumn{4}{c}{{\small $\mathrm{L_2}$}}  \\
 		\hline
		 \multicolumn{2}{c|}{} &{\small $\mathrm{Reg_{sr}^{25}}$} &{\small $\mathrm{Reg_{sr}^{50}}$} &{\small $\mathrm{Reg_{sr}^{75}}$} &{\small $\mathrm{Reg_{sr}^{100}}$}\\
 		\hline
		  \multicolumn{2}{c|}{HardNetF} &\makecell{10.39\%} &\makecell{11.04\%} &\makecell{11.04\%} &\makecell{11.04\%} \\
	     \hline
 		\multicolumn{2}{c|}{SosNetF} &\makecell{5.19\%} &\makecell{5.84\%} &\makecell{5.84\% }&\makecell{5.84\% }\\
 		\hline
 		\multicolumn{2}{c|}{HyNetF} &\makecell{5.84\%} &\makecell{5.84\%} &\makecell{5.84\%} &\makecell{7.14\% }\\
 		\hline
 		\multicolumn{2}{c|}{SDGMNetF} &\makecell{5.84\%} &\makecell{6.49\%} &\makecell{6.49\%} &\makecell{6.49\% }\\
 		\hline
 		\multicolumn{2}{c|}{Ours}  &\makecell{\textbf{69.48\%}} 
            &\makecell{\textbf{79.87\%}}&\makecell{\textbf{79.87\%}}&\makecell{\textbf{79.87\%}}\\
 		\hline
 	\end{tabular}\label{tab_result_compare}
 \end{table}
We compare our approach quantitatively with other feature-based approaches. 
Fig. \ref{box_compare} shows the MEEs of the different methods on the four levels of the testing set, and Table \ref{tab_result_compare} lists the registration success ratios of the different approaches, demonstrating that our method consistently achieves higher success ratios. The number of correctly matched point pairs is crucial for feature-based method performance. As the Region of Search (ROS) increases, feature-based methods require extracting more keypoints from optical images, increasing both memory and time costs. As depicted in Fig. \ref{keypoints_compare}, our Grid-Reg uses significantly fewer grid points compared to the feature-based methods, particularly as the ROS increases. For example, in the ID 4 testing case, where all methods achieve successful registration, Grid-Reg attains lower MEE while using fewer grid points.
\begin{figure}[!htbp]
	\centering
	\includegraphics[width=0.9\linewidth]{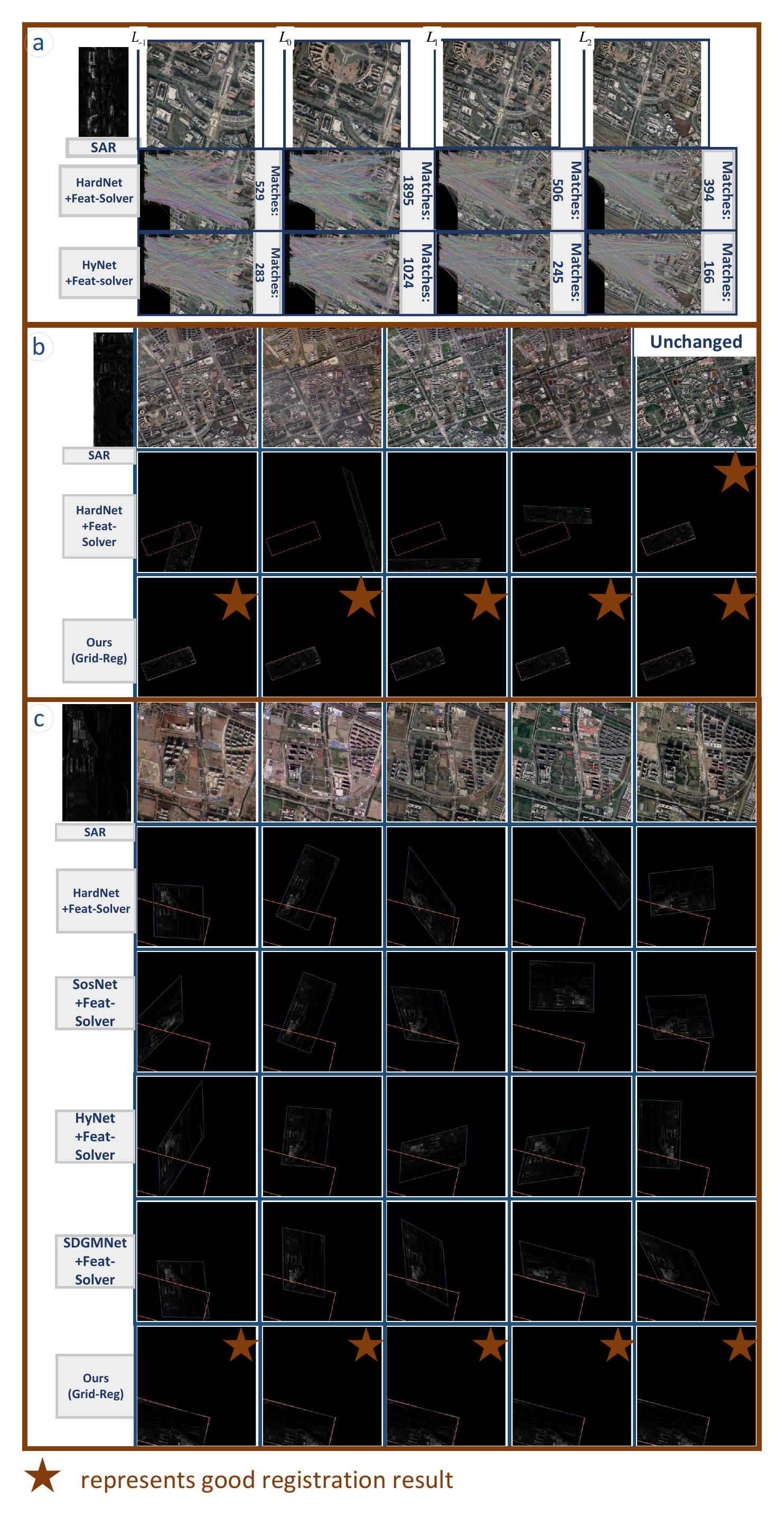}
	\centering
	\caption{The results of our method and other feature-based methods. (a) test ID 8, (b) test ID 6 at $\mathrm{L}_\mathrm{2}$, (c) test ID 5 at $\mathrm{L}_\mathrm{-1}$.}\label{Compare_5_L1}
\end{figure} 
As shown in Fig. \ref{box_compare}, taking test ID 8 as an example, compared to {\small $\mathrm{L_0}$}, the performance of feature-based methods notably decreases at $\mathrm{L}_\mathrm{-1}$ and $\mathrm{L}_\mathrm{2}$ levels. Fig. \ref{Compare_5_L1}(a) illustrates that this decline results from a reduction in the number of correctly matched point pairs as registration difficulty increases. The decrease in correct matching point pairs with increasing difficulty stems from two main reasons. Firstly, as ROS expands, the likelihood of similar structures appearing increases, particularly in complex urban scenes with numerous buildings sharing similar structures. Secondly, as ROA shrinks, the availability of features for correct matching diminishes. As shown in Fig. \ref{Compare_5_L1}(c), in regions lacking sufficient salient features, our Grid-Reg outperforms feature-based approaches. This advantage of our method becomes more apparent because our method overcomes the issues caused by non-repeatable keypoints. Furthermore, our Grid-Reg employs a novel global and robust matching loss to mitigate the impact of numerous false correspondences. In addition, the changing areas between images affect image registration performance. As shown in the Fig. \ref{Compare_5_L1}(b), feature-based methods fail when buildings change. Our method achieves better registration performance, unaffected by these changing areas.
\subsection{Ablation Analysis of Grid-Solver}
\begin{table}[!htbp]
 	\caption{Comparing the success rate of HSCMLNet with Feat-Solver(F) and Grid-Solver(G).}
	\centering
 	\begin{tabular}{c|c|c|c|c|c}
 		\hline
 		\multicolumn{2}{c|}{Method} &\multicolumn{4}{c}{$\mathrm{L}_\mathrm{-1}$}  \\
 		\hline
 		\multicolumn{2}{c|}{} &{\small $\mathrm{Reg_{sr}^{25}}$} &{\small $\mathrm{Reg_{sr}^{50}}$} &{\small $\mathrm{Reg_{sr}^{50}}$} &{\small $\mathrm{Reg_{sr}^{100}}$}\\
 		\hline
            \multicolumn{2}{c|}{HSCMLNetF}   &\makecell{14.29\%} &\makecell{21.43\%}&\makecell{24.03\%}&\makecell{26.62\%}\\
            \hline
 		\multicolumn{2}{c|}{HSCMLNetG}  &\makecell{\textbf{65.58\%}} 
            &\makecell{\textbf{75.32\%}}&\makecell{\textbf{76.62\%}}&\makecell{\textbf{77.92\%}}\\
 		\hline
 		\multicolumn{2}{c|}{} &\multicolumn{4}{c}{{\small $\mathrm{L_0}$}}  \\
 		\hline
 		\multicolumn{2}{c|}{} &{\small $\mathrm{Reg_{sr}^{25}}$} &{\small $\mathrm{Reg_{sr}^{50}}$} &{\small $\mathrm{Reg_{sr}^{75}}$} &{\small $\mathrm{Reg_{sr}^{100}}$}\\
 		\hline
            \multicolumn{2}{c|}{HSCMLNetF}&\makecell{29.22\%} &\makecell{36.36\%} &\makecell{36.36\%}&\makecell{39.61\%}\\
            \hline
 		\multicolumn{2}{c|}{HSCMLNetG}  &\makecell{\textbf{80.52\%}} &\makecell{\textbf{85.71\%}}&\makecell{\textbf{85.71\%}}&\makecell{\textbf{87.01\%}}\\
 		\hline
 		\multicolumn{2}{c|}{} &\multicolumn{4}{c}{{\small $\mathrm{L_1}$}}  \\
 		\hline
 		\multicolumn{2}{c|}{} &{\small $\mathrm{Reg_{sr}^{25}}$} &{\small $\mathrm{Reg_{sr}^{50}}$} &{\small $\mathrm{Reg_{sr}^{75}}$} &{\small $\mathrm{Reg_{sr}^{100}}$}\\
 		\hline
    	\multicolumn{2}{c|}{HSCMLNetF}&\makecell{24.03\%} &\makecell{28.57\%} &\makecell{31.17\%}&\makecell{31.17\% }\\
    	\hline
 		\multicolumn{2}{c|}{HSCMLNetG}  &\makecell{\textbf{78.57\%}} &\makecell{\textbf{83.12\%}}&\makecell{\textbf{84.42\%}}&\makecell{\textbf{84.42\%}}\\
 		\hline
		 \multicolumn{2}{c|}{} &\multicolumn{4}{c}{{\small $\mathrm{L_2}$}}  \\
 		\hline
		 \multicolumn{2}{c|}{} &{\small $\mathrm{Reg_{sr}^{25}}$} &{\small $\mathrm{Reg_{sr}^{50}}$} &{\small $\mathrm{Reg_{sr}^{75}}$} &{\small $\mathrm{Reg_{sr}^{100}}$}\\
 		\hline
		  \multicolumn{2}{c|}{HSCMLNetF} &\makecell{22.73\%} &\makecell{28.57\%} &\makecell{29.87\%} &\makecell{30.52\%} \\
	     \hline
 		\multicolumn{2}{c|}{HSCMLNetG}  &\makecell{\textbf{69.48\%}} 
            &\makecell{\textbf{79.87\%}}&\makecell{\textbf{79.87\%}}&\makecell{\textbf{79.87\%}}\\
 		\hline
 	\end{tabular}\label{tab_result_REG}
 \end{table}
Table. \ref{tab_result_REG} quantitatively compares the performance of Feat-Solver and Grid-Solver on the test sets of different difficulty levels using the same feature extractor HSCMLNet. Compared with Feat-Solver, Grid-Solver leads to a significantly higher registration success rate, especially when the ROA is reduced. 
\subsection{Ablation Analysis of HSCMLNet}
\begin{figure}[!htbp]
	\centering
	\includegraphics[width=0.9\linewidth]{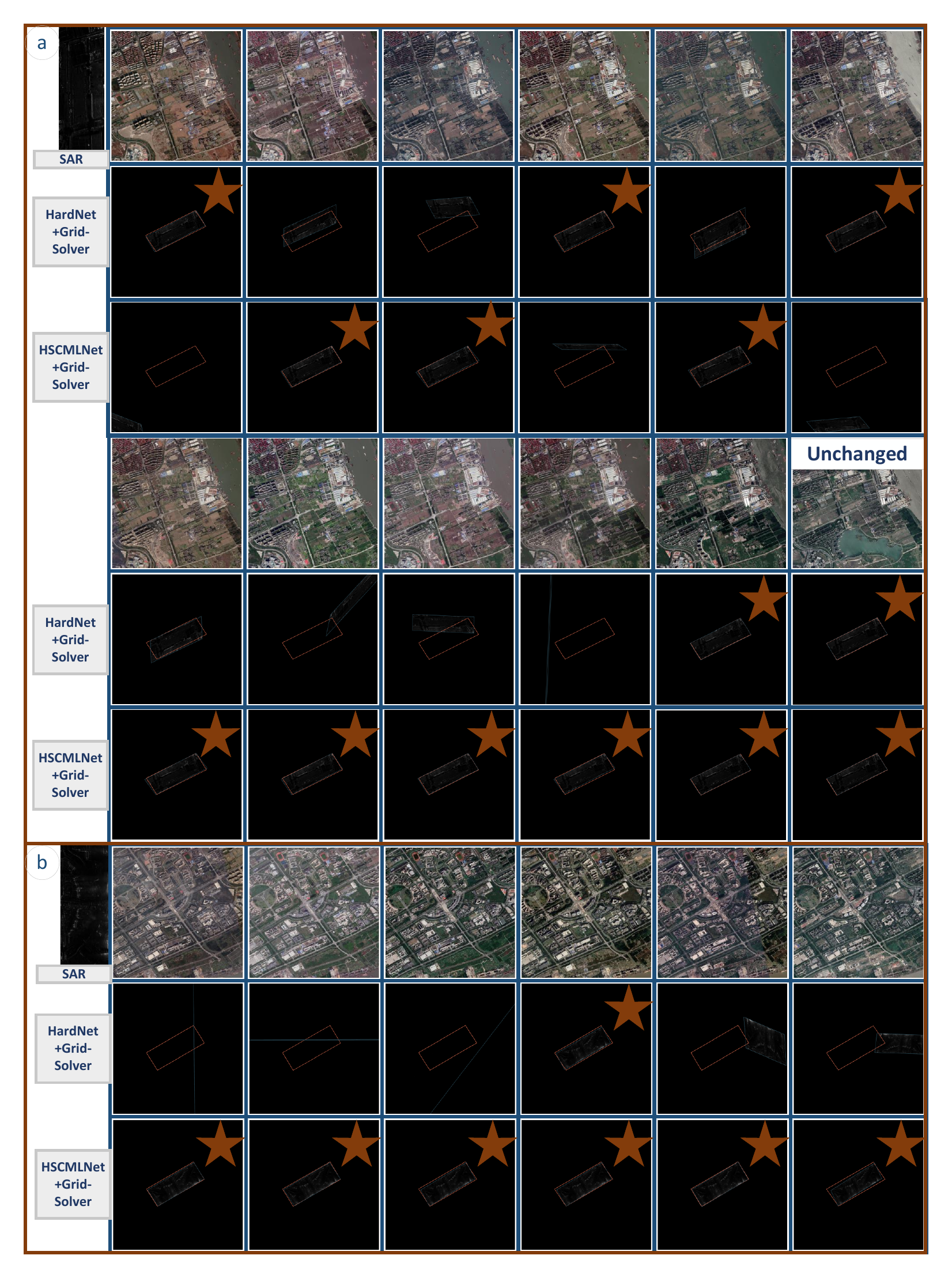}
	\centering
	\caption{The results of HardNet+Grid-Solver and HSCMLNet+Grid-Solver. (a) test ID 9 at $\mathrm{L}_\mathrm{2}$,(b)  test ID 11 at $\mathrm{L}_\mathrm{2}$.}\label{Compare_9_L4}
\end{figure}
\begin{table}[!htbp]
 	\caption{Compare the success rates of the different descriptor methods with the same Grid-Solver(G).}
	\centering
 	\begin{tabular}{c|c|c|c|c|c}
 		\hline
 		\multicolumn{2}{c|}{Method} &\multicolumn{4}{c}{$\mathrm{L}_\mathrm{-1}$}  \\
 		\hline
 		\multicolumn{2}{c|}{} &{\small $\mathrm{Reg_{sr}^{25}}$} &{\small $\mathrm{Reg_{sr}^{50}}$} &{\small $\mathrm{Reg_{sr}^{50}}$} &{\small $\mathrm{Reg_{sr}^{100}}$}\\
 		\hline
            \multicolumn{2}{c|}{HardNetG}   &\makecell{32.47\%} &\makecell{33.77\%}&\makecell{33.77\%}&\makecell{33.77\%}\\
            \hline
     	\multicolumn{2}{c|}{SosNetG}  &\makecell{24.68\%} &\makecell{24.68\%} &\makecell{24.68\%} 
            &\makecell{24.68\%} \\
     	\hline
		  \multicolumn{2}{c|}{HyNetG}&\makecell{28.57\%} &\makecell{32.47\%} &\makecell{32.47\%} 
            &\makecell{32.47\%} \\
 		\hline
 		\multicolumn{2}{c|}{SDGMNetG}&\makecell{20.78\%} &\makecell{21.43\%} 
            &\makecell{21.43\%}&\makecell{21.43\%}\\
 		\hline
 		\multicolumn{2}{c|}{HSCMLNetG}  &\makecell{\textbf{65.58\%}} 
            &\makecell{\textbf{75.32\%}}&\makecell{\textbf{76.62\%}}&\makecell{\textbf{77.92\%}}\\
 		\hline
 		\multicolumn{2}{c|}{} &\multicolumn{4}{c}{{\small $\mathrm{L_0}$}}  \\
 		\hline
 		\multicolumn{2}{c|}{} &{\small $\mathrm{Reg_{sr}^{25}}$} &{\small $\mathrm{Reg_{sr}^{50}}$} &{\small $\mathrm{Reg_{sr}^{75}}$} &{\small $\mathrm{Reg_{sr}^{100}}$}\\
 		\hline
            \multicolumn{2}{c|}{HardNetG}&\makecell{42.86\%} &\makecell{45.45\%} &\makecell{45.45\%}&\makecell{45.45\%}\\
            \hline
 		\multicolumn{2}{c|}{SosNetG}&\makecell{31.82\%} &\makecell{32.47\%} &\makecell{32.47\%}  &\makecell{32.47\%}\\
 		\hline
 		\multicolumn{2}{c|}{HyNetG} &\makecell{40.91\%} &\makecell{42.21\%} &\makecell{42.21\%} &\makecell{42.21\%}\\
 		\hline
 		\multicolumn{2}{c|}{SDGMNetG}&\makecell{31.17\%} &\makecell{31.17\%} &\makecell{31.17\%}&\makecell{31.17\%}\\
 		\hline
 		\multicolumn{2}{c|}{HSCMLNet
G}  &\makecell{\textbf{80.52\%}} &\makecell{\textbf{85.71\%}}&\makecell{\textbf{85.71\%}}&\makecell{\textbf{87.01\%}}\\
 		\hline
 		\multicolumn{2}{c|}{} &\multicolumn{4}{c}{{\small $\mathrm{L_1}$}}  \\
 		\hline
 		\multicolumn{2}{c|}{} &{\small $\mathrm{Reg_{sr}^{25}}$} &{\small $\mathrm{Reg_{sr}^{50}}$} &{\small $\mathrm{Reg_{sr}^{75}}$} &{\small $\mathrm{Reg_{sr}^{100}}$}\\
 		\hline
    	\multicolumn{2}{c|}{HardNetG}&\makecell{38.96\%} &\makecell{38.96\%} &\makecell{38.96\%}&\makecell{38.96\% }\\
    	\hline
 		\multicolumn{2}{c|}{SosNetG} &\makecell{27.92\%} &\makecell{28.57\%} &\makecell{28.57\%} &\makecell{28.57\% }\\
 		\hline
 		\multicolumn{2}{c|}{HyNetG} &\makecell{36.36\%} &\makecell{38.31\%} &\makecell{38.31\% }&\makecell{38.31\%}\\
 		\hline
 		\multicolumn{2}{c|}{SDGMNetG} &\makecell{26.62\%} &\makecell{26.62\%} &\makecell{26.62\%} &\makecell{26.62\% }\\
 		\hline
 		\multicolumn{2}{c|}{HSCMLNetG}  &\makecell{\textbf{78.57\%}} &\makecell{\textbf{83.12\%}}&\makecell{\textbf{84.42\%}}&\makecell{\textbf{84.42\%}}\\
 		\hline
		 \multicolumn{2}{c|}{} &\multicolumn{4}{c}{{\small $\mathrm{L_2}$}}  \\
 		\hline
		 \multicolumn{2}{c|}{} &{\small $\mathrm{Reg_{sr}^{25}}$} &{\small $\mathrm{Reg_{sr}^{50}}$} &{\small $\mathrm{Reg_{sr}^{75}}$} &{\small $\mathrm{Reg_{sr}^{100}}$}\\
 		\hline
		  \multicolumn{2}{c|}{HardNetG} &\makecell{32.47\%} &\makecell{33.77\%} &\makecell{33.77\%} &\makecell{34.42\%} \\
	     \hline
 		\multicolumn{2}{c|}{SosNetG} &\makecell{22.73\%} &\makecell{22.73\%} &\makecell{22.73\% }&\makecell{22.73\% }\\
 		\hline
 		\multicolumn{2}{c|}{HyNetG} &\makecell{27.27\%} &\makecell{30.52\%} &\makecell{30.52\%} &\makecell{31.17\% }\\
 		\hline
 		\multicolumn{2}{c|}{SDGMNetG} &\makecell{22.72\%} &\makecell{22.72\%} &\makecell{22.72\%} &\makecell{22.72\% }\\
 		\hline
 		\multicolumn{2}{c|}{HSCMLNetG}  &\makecell{\textbf{69.48\%}} 
            &\makecell{\textbf{79.87\%}}&\makecell{\textbf{79.87\%}}&\makecell{\textbf{79.87\%}}\\
 		\hline
 	\end{tabular}\label{tab_result_A1}
 \end{table}
In Fig.~\ref{box_compare} and Table.~\ref{tab_result_A1}, we compare HSCMLNet with other descriptor learning networks quantitatively, showing the superiority of our HSCMLNet over other networks when using the same Grid-Solver. Comparing Table. ~\ref{tab_result_compare} and \ref{tab_result_A1}, our Grid-Solver improves the performance of other local descriptor methods compared to Feat-Solver. Fig. ~\ref{box_compare} shows HSCMLNet achieving better registration results, especially for the SAR image of poor quality (e.g., test ID 0, 11). The flight instability introduces noise, causing incomplete and blurred buildings in the SAR image, reducing discriminative regions and making it be difficult to measure the similarity between patches. Fig.~\ref{box_compare} also indicates that HardNet + Grid-Solver outperforms other local descriptor methods, except HSCMLNet. Fig.~\ref{Compare_9_L4}(b) demonstrates that the HSCMLNet works better even for a SAR image of poor quality than other methods, highlighting its robustness. Temporal differences between SAR and optical images introduce changing objects, significantly impacting registration. Fig. ~\ref{Compare_9_L4}(a) demonstrates that HSCMLNet consistently outperforms other methods in scenarios containing numerous changing regions between SAR and optical images, highlighting its robustness against interference introduced by such changes.
 \begin{figure}[!htbp]
	\centering
	\includegraphics[width=\linewidth]{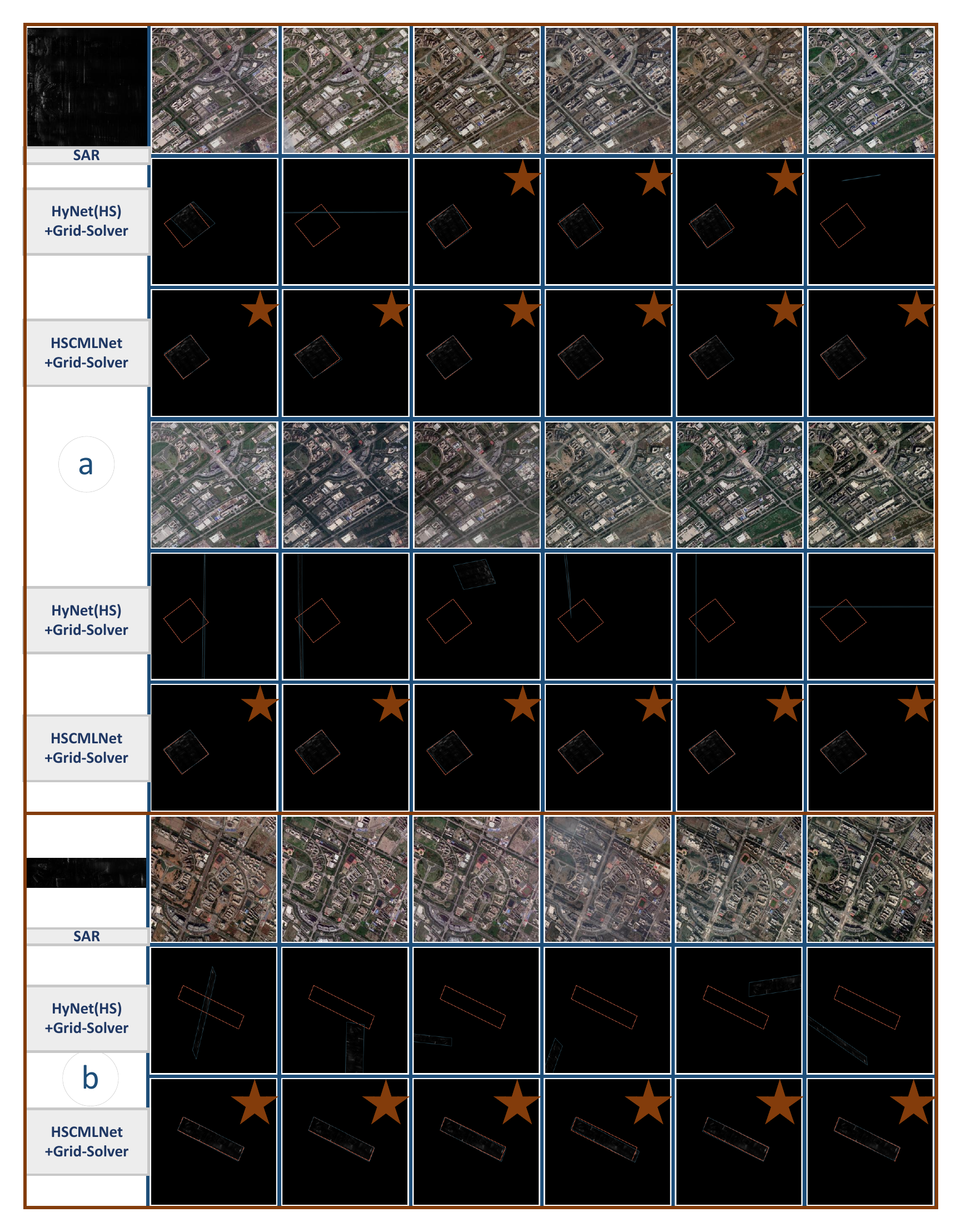}
	\centering
	\caption{The results of HSCMLNet+Grid-Solver and HyNet(HS)+Grid-Solver. (a)test ID 0 at {\small $\mathrm{L}_\mathrm{2}$},(a)test ID 3 at {\small $\mathrm{L}_\mathrm{2}$}.}\label{Compare_0_L4}
\end{figure} 
\subsection{Ablation Analysis of CMLModule and Loss function }
\begin{table}[!htbp]
 \caption{Comparing the performance of our method with other methods, which utilize HSModule and Grid-Solver(G).}
	\centering
 	\begin{tabular}{c|c|c|c|c|c}
 		\hline
 		\multicolumn{2}{c|}{Method} &\multicolumn{4}{c}{$\mathrm{L}_\mathrm{-1}$}  \\
 		\hline
 		\multicolumn{2}{c|}{} &{\small $\mathrm{Reg_{sr}^{25}}$} &{\small $\mathrm{Reg_{sr}^{50}}$} &{\small $\mathrm{Reg_{sr}^{75}}$} &{\small $\mathrm{Reg_{sr}^{100}}$}\\
 		\hline
            \multicolumn{2}{c|}{HardNet(HS)G}   &\makecell{30.52\%} &\makecell{31.82\%}&\makecell{31.82\%}&\makecell{32.47\%}\\
            \hline
     	\multicolumn{2}{c|}{SosNet(HS)G}  &\makecell{35.06\%} &\makecell{35.06\%} &\makecell{35.71\%} 
            &\makecell{36.36\%} \\
     	\hline
		  \multicolumn{2}{c|}{HyNet(HS)G}&\makecell{44.16\%} &\makecell{48.70\%} &\makecell{49.35\%} 
            &\makecell{49.35\%} \\
 		\hline
 		\multicolumn{2}{c|}{SDGMNet(HS)G}&\makecell{27.92\%} &\makecell{29.22\%} 
            &\makecell{29.22\%}&\makecell{29.22\%}\\
 		\hline
 		\multicolumn{2}{c|}{HSCMLNetG}  &\makecell{\textbf{65.58\%}} 
            &\makecell{\textbf{75.32\%}}&\makecell{\textbf{76.62\%}}&\makecell{\textbf{77.92\%}}\\
 		\hline
 		\multicolumn{2}{c|}{} &\multicolumn{4}{c}{{\small $\mathrm{L_0}$}}  \\
 		\hline
 		\multicolumn{2}{c|}{} &{\small $\mathrm{Reg_{sr}^{25}}$} &{\small $\mathrm{Reg_{sr}^{50}}$} &{\small $\mathrm{Reg_{sr}^{75}}$} &{\small $\mathrm{Reg_{sr}^{100}}$}\\
 		\hline
            \multicolumn{2}{c|}{HardNet(HS)G}&\makecell{38.31\%} &\makecell{38.31\%} &\makecell{38.31\%}&\makecell{38.31\%}\\
            \hline
 		\multicolumn{2}{c|}{SosNet(HS)G}&\makecell{51.95\%} &\makecell{53.90\%} &\makecell{53.90\%}  &\makecell{53.90\%}\\
 		\hline
 		\multicolumn{2}{c|}{HyNet(HS)G} &\makecell{63.64\%} &\makecell{67.53\%} &\makecell{67.53\%} &\makecell{67.53\%}\\
 		\hline
 		\multicolumn{2}{c|}{SDGMNet(HS)G}&\makecell{40.91\%} &\makecell{42.21\%} &\makecell{42.21\%}&\makecell{42.21\%}\\
 		\hline
 		\multicolumn{2}{c|}{HSCMLNetG}  &\makecell{\textbf{80.52\%}} &\makecell{\textbf{85.71\%}}&\makecell{\textbf{85.71\%}}&\makecell{\textbf{87.01\%}}\\
 		\hline
 		\multicolumn{2}{c|}{} &\multicolumn{4}{c}{{\small $\mathrm{L_1}$}}  \\
 		\hline
 		\multicolumn{2}{c|}{} &{\small $\mathrm{Reg_{sr}^{25}}$} &{\small $\mathrm{Reg_{sr}^{50}}$} &{\small $\mathrm{Reg_{sr}^{75}}$} &{\small $\mathrm{Reg_{sr}^{100}}$}\\
 		\hline
    	\multicolumn{2}{c|}{HardNet(HS)G}&\makecell{38.31\%} &\makecell{38.31\%} &\makecell{38.31\%}&\makecell{38.31\% }\\
    	\hline
 		\multicolumn{2}{c|}{SosNet(HS)G} &\makecell{44.81\%} &\makecell{47.40\%} &\makecell{48.05\%} &\makecell{48.05\% }\\
 		\hline
 		\multicolumn{2}{c|}{HyNet(HS)G} &\makecell{51.95\%} &\makecell{57.14\%} &\makecell{57.14\% }&\makecell{57.14\%}\\
 		\hline
 		\multicolumn{2}{c|}{SDGMNet(HS)G} &\makecell{38.96\%} &\makecell{39.61\%} &\makecell{40.26\%} &\makecell{40.26\% }\\
 		\hline
 		\multicolumn{2}{c|}{HSCMLNetG}  &\makecell{\textbf{78.57\%}} &\makecell{\textbf{83.12\%}}&\makecell{\textbf{84.42\%}}&\makecell{\textbf{84.42\%}}\\
 		\hline
		 \multicolumn{2}{c|}{} &\multicolumn{4}{c}{{\small $\mathrm{L_2}$}}  \\
 		\hline
		 \multicolumn{2}{c|}{} &{\small $\mathrm{Reg_{sr}^{25}}$} &{\small $\mathrm{Reg_{sr}^{50}}$} &{\small $\mathrm{Reg_{sr}^{75}}$} &{\small $\mathrm{Reg_{sr}^{100}}$}\\
 		\hline
		  \multicolumn{2}{c|}{HardNet(HS)G} &\makecell{35.71\%} &\makecell{36.36\%} &\makecell{36.36\%} &\makecell{36.36\%} \\
	     \hline
 		\multicolumn{2}{c|}{SosNet(HS)G} &\makecell{40.91\%} &\makecell{42.86\%} &\makecell{42.86\% }&\makecell{43.51\% }\\
 		\hline
 		\multicolumn{2}{c|}{HyNet(HS)G} &\makecell{40.26\%} &\makecell{44.16\%} &\makecell{44.16\%} &\makecell{44.81\% }\\
 		\hline
 		\multicolumn{2}{c|}{SDGMNet(HS)G} &\makecell{40.91\%} &\makecell{41.56\%} &\makecell{41.56\%} &\makecell{41.56\% }\\
 		\hline
 		\multicolumn{2}{c|}{HSCMLNetG}  &\makecell{\textbf{69.48\%}} 
            &\makecell{\textbf{79.87\%}}&\makecell{\textbf{79.87\%}}&\makecell{\textbf{79.87\%}}\\
 		\hline
 	\end{tabular}\label{tab_result_A2}
 \end{table}
To validate our CMLModule and loss functions, we replace the feature extraction modules of HardNet, SosNet, HyNet, and SDGMNet with HSModule. The orange parts (HardModule, SosModule, HyModule, SDGMModule) are combined with HSModule to be new architectures: HardNet(HS), SosNet(HS), HyNet(HS), and SDGMNet(HS). Table.~\ref{tab_result_A2} shows that our method outperforms HardNet(HS), SosNet(HS), HyNet(HS), and SDGMNet(HS), especially at levels {\small $\mathrm{L}_\mathrm{-1}$} and {\small $\mathrm{L_2}$}. Fig.~\ref{box_compare} shows the performance of the methods across test image subsets. Notably, for test ID 0, 1, 2, 3, 7, 9, and 11, our CMLModule outperforms other networks. In scenarios with significant SAR image noise (Fig.~\ref{Compare_0_L4}(a)), our method successfully alignment SAR with optical images despite obscured objects. Landmark buildings are critical for registration. In the scenarios without landmark buildings (Fig.~\ref{Compare_0_L4}(b)), our method shows better resistance to optical image interference from similar structured buildings compared to HyNet (HS) + Grid-Solver.
 \begin{figure}[!htbp]
	\centering
	\includegraphics[width=\linewidth]{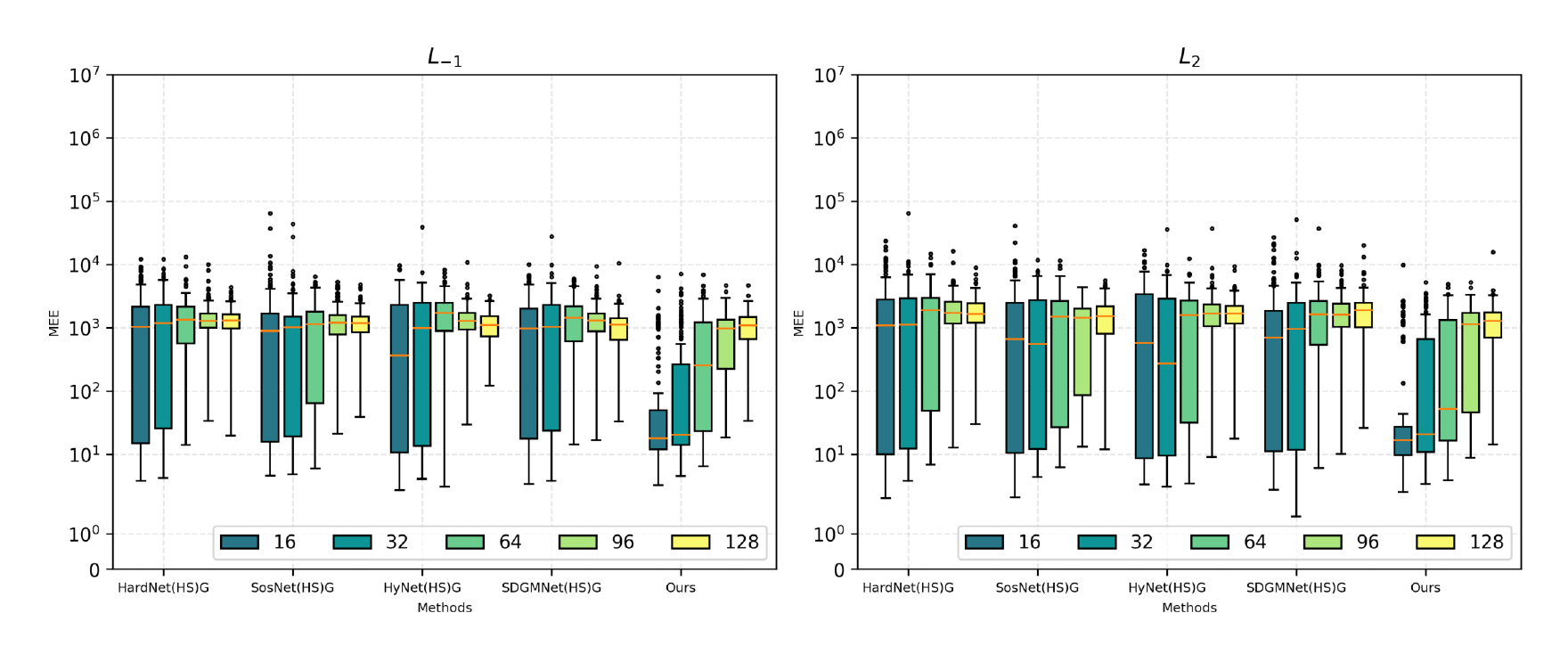}
	\centering
	\caption{When {\small $\mathrm{Step=16,32,64,96,128}$}, MEE of HardNet(HS), SosNet(HS), HyNet(HS), and SDGMNet(HS) with Grid-Solver(G) at $\mathrm{L}_\mathrm{-1}$ and $\mathrm{L}_\mathrm{2}$.}\label{Step}
\end{figure}

\begin{figure}[!htbp]
	\centering
	\includegraphics[width=\linewidth]{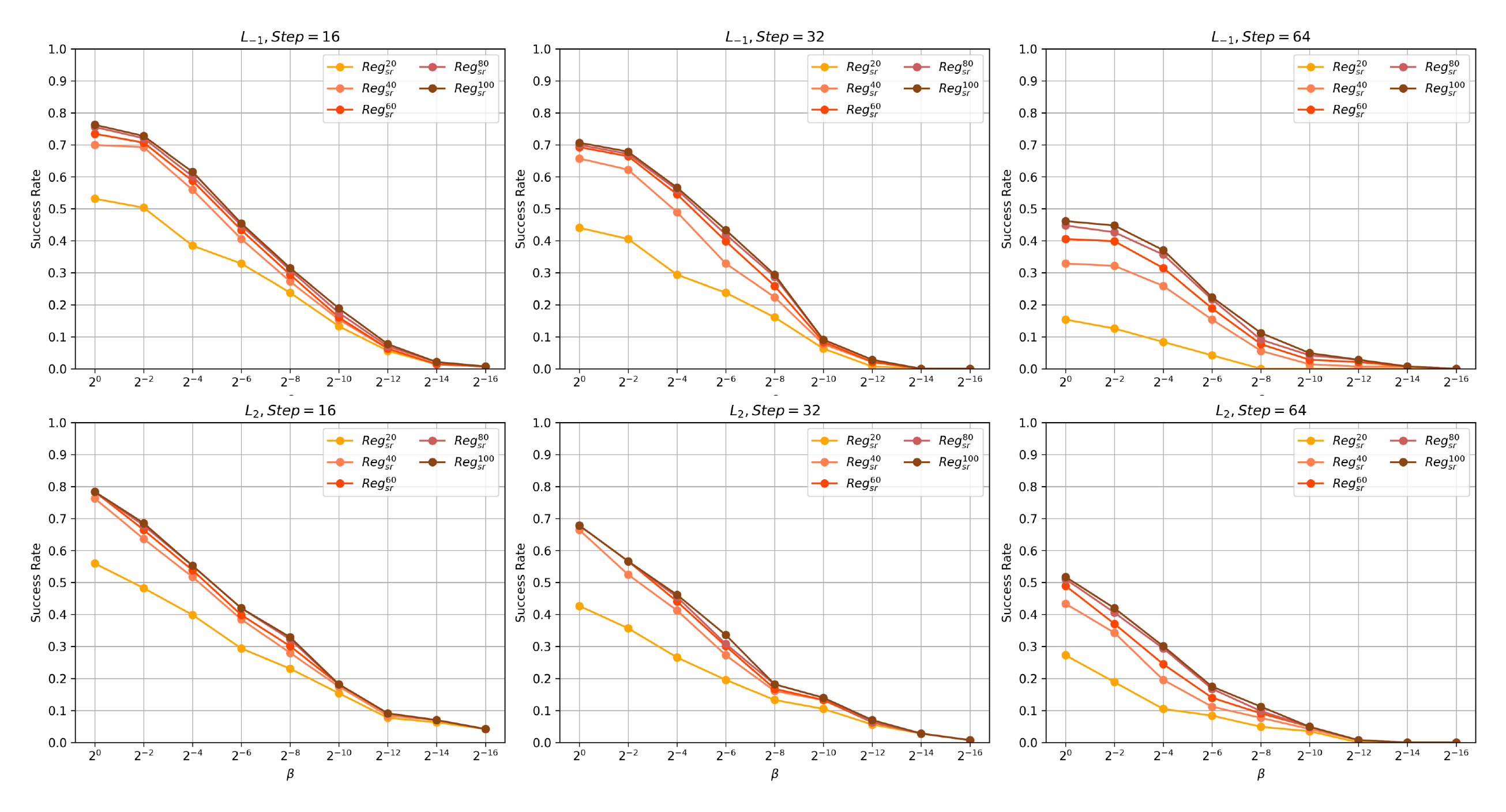}
	\centering
	\caption{The registration success rate of our method varies with Step and number of iterations at {\small $\mathrm{L}_\mathrm{-1}$} and $\mathrm{L}_\mathrm{2}$}\label{ITERSTEP}
\end{figure}
\subsection{Discussion of Time and Memory Consumption}
The step of the sliding window {\small $\mathrm{Step}$} and the number of iterations  {\small $\mathrm{N}_{\mathrm{iter}}^\mathrm{outer}$} for transformation estimation are the key hyperparameters related to registration performance, computation time, and memory consumption.
\subsubsection{The Variation of {\small $\mathrm{Step}$}}
We compare our method with other methods that use HSModule under {\small $\mathrm{Step=16,32,64,128}$}. 
Fig. \ref{Step} shows that as the step size increases, the MEE of various methods also increases.
But compared with others, our method can retain better results for step sizes up to 96. Particularly, when the step size is 64 or less, our method significantly outperforms other methods, demonstrating its superior balance between 
accuracy and computational efficiency.
\subsubsection{The combined variation of {\small $\mathrm{Step}$} and {\small $\mathrm{N}_{\mathrm{iter}}^\mathrm{outer}$}}
Fig.~\ref{ITERSTEP} illustrates the effect of reducing the iteration number on the performance of our method for step sizes of 16, 32, and 64. When {\small $\beta \geq 0.1$}, decreasing $\beta$ results in only a slightly decline of the registration success rate of our approach. This approach exhibits more sensitivity to the iteration numbers at {\small $\mathrm{L_2}$} than at {\small $\mathrm{L}_\mathrm{-1}$}, due to the larger search range required for accurate transformation in higher RSO conditions. For {\small $\beta \in [2^0, 2^{-2}]$}, decreasing $\beta$ moderately reduces the registration success rate of our method, indicating a balance between accuracy and time efficiency within this range. Different step sizes show varying impacts of {\small $\beta$} reduction on success rate. At {\small $\mathrm{Step = 16}$, $\mathrm{Reg_{sr}^{20}}$} remains above 20\% even at {\small $\beta = 2^{-8}$}. Conversely, at {\small $\mathrm{Step = 64}$}, {\small $\mathrm{Reg_{sr}^{20}}$} drops below 10\% at {\small $\beta = 2^{-2}$}. 


\section{Conclusion}
In this paper, we present a novel grid-based multimodal image registration framework, Grid-Reg, for large-scale SAR and optical imagery registration with significant geometric differences and modality gaps. Grid-Reg integrates HSCMLNet and Grid-Solver. The HSCMLNet, a hybrid Siamese correction metric learning network, is designed to extract robust patch-level descriptors from both SAR and optical image patches. The HSCMLNet comprises the HSModule and the CMLModule based on EUBVs representation. To enhance modality-invariant descriptor learning, we introduce a manifold-aware cross-modality EUBVs reconstruction loss and a joint-modality EUBVs reconstruction loss, which are jointly optimized with a contrastive loss. This combination enables HSCMLNet to extract robust patch-level descriptors against substantial geometric, radiometric, and temporal variations, as well as noise. To estimate the transformation parameters, we proposed Grid-Solver, a grid-based optimal parameter search strategy that minimizes the proposed matching loss, thereby avoiding the fragility of keypoint detection in cross-modal settings.

We evaluated Grid-Reg on a newly curated large-scale, SAR and optical image registration dataset, performing extensive quantitative and qualitative comparisons against state-of-the-art approaches. Experimental results demonstrate that our method consistently achieves superior registration accuracy, particularly in challenging scenarios with extensive temporal changes, geometric variations, and noise.

\bibliographystyle{ieeetr}
\bibliography{bibtex}
\end{document}